\documentclass[11pt,a4paper]{JHEP3}

\usepackage{amsmath,graphicx,amsfonts, mathrsfs, amssymb,cite}

\newcommand{\be}{\begin{equation}}
\newcommand{\ee}{\end{equation}}
\newcommand{\ba}{\begin{eqnarray}}
\newcommand{\ea}{\end{eqnarray}}
\newcommand{\no}{\nonumber \\}
\newcommand{\gsim}{\mathrel{\hbox{\rlap{\lower.55ex \hbox {$\sim$}}
                   \kern-.3em \raise.4ex \hbox{$>$}}}}
\newcommand{\lsim}{\mathrel{\hbox{\rlap{\lower.55ex \hbox {$\sim$}}
                   \kern-.3em \raise.4ex \hbox{$<$}}}}

\def\be{\begin{eqnarray}}
\def\ee{\end{eqnarray}}

\def\del{{\partial}}

\def\roughly#1{\mathrel{\raise.3ex\hbox{$#1$\kern-.75em%
\lower1ex\hbox{$\sim$}}}}
\def\lsim{\roughly<}
\def\gsim{\roughly>}

\def\vx{{\vec x}}

\def\({\left(}
\def\){\right)}
\def\[{\left[}
\def\]{\right]}

\def\lsim{\mathrel{\rlap{\lower3pt\hbox{\hskip1pt$\sim$}}
     \raise1pt\hbox{$<$}}} 
\def\gsim{\mathrel{\rlap{\lower3pt\hbox{\hskip1pt$\sim$}}
     \raise1pt\hbox{$>$}}} 
\def\N{{\mathcal{N}}}
\def\jt{\langle J^t \rangle}
\def\Om{{\mathcal{O}_M}}
\def\Omv{\langle {\mathcal{O}_M} \rangle}

\def\dlt{\delta}
\def\eps{\epsilon}
\def\omg{\omega}
\def\Omg{\Omega}

\def\ap{\alpha'{}}
\def\wg{\wedge}
\def\Y{{\cal Y}}
\def\veps{\varepsilon}

\def\bN{\bar{N}}
\def\brho{\bar{\rho}}
\def\by{\bar{y}}
\def\bomg{\bar{\omg}}
\def\bk{\bar{k}}
\def\a{\alpha}
\def\b{\beta}
\def\g{\gamma}

\title{\LARGE On Stability and Transport of Cold Holographic Matter}

\author{Martin Ammon,$^1$\footnotemark[1]\, Johanna Erdmenger,$^2$\footnotemark[2]\, Shu Lin,$^2$\footnotemark[2]\, Steffen M\"uller,$^2$\footnotemark[2]\, Andy O'Bannon$^3$\footnotemark[3]\, and Jonathan P. Shock$^2$\footnotemark[2]\,
\\
$^1$Department of Physics and Astronomy, University of California \\ Los Angeles, CA 90095, United States
\\
\\
$^2$Max-Planck-Institut f\"{u}r Physik (Werner-Heisenberg-Institut) \\ F\"{o}hringer Ring 6, 80805 M\"{u}nchen, Germany
\\
\\
$^3$Department of Applied Mathematics and Theoretical Physics\\ University of Cambridge\\ Cambridge CB3 0WA, United Kingdom}

\footnotetext[1]{E-mail address: \email{ammon@physics.ucla.edu}}
\footnotetext[2]{E-mail addresses: \email{jke,slin,smueller,jonshock@mppmu.mpg.de}}
\footnotetext[3]{E-mail address: \email{A.OBannon@damtp.cam.ac.uk}}

\abstract{We use gauge-gravity duality to study the stability of zero-temperature, finite baryon density states of $\N=4$ supersymmetric $SU(N_c)$ Yang-Mills theory coupled to a single massive fundamental-representation $\N=2$ hypermultiplet in the large-$N_c$ and large-coupling limits. In particular, we study the spectrum of mesons. The dual description is a probe D7-brane in anti-de Sitter space with a particular configuration of worldvolume fields. The meson spectrum is dual to the spectrum of fluctuations of worldvolume fields about that configuration. We use a combination of analytical and numerical techniques to compute the spectrum, including a special numerical technique designed to deal with singular points in the fluctuations' equations of motion. Despite circumstantial evidence that the system might be unstable, such as a finite entropy density at zero temperature and the existence of instabilities in similar theories, we find no evidence of any instabilities, at least for the ranges of frequency and momenta that we consider. We discover a pole on the imaginary frequency axis in a scalar meson two-point function, similar to the diffusive mode in the two-point function of a conserved charge.}


\preprint{DAMTP-2011-60\\MPP-2011-96}

\begin{document}

\maketitle

\section{Introduction}

Many real physical systems involve a finite chemical potential and strongly-interacting degrees of freedom. Examples include Quantum Chromodynamics (QCD) with a finite baryon number chemical potential and many condensed matter systems. 

Such systems often exhibit unusual transport properties. For example, the quark-gluon plasma (QGP) created at the Relativistic Heavy-Ion Collider (RHIC) appears to have the lowest viscosity of any known substance 
\cite{Shuryak:2003xe}. Many condensed matter systems, such as  the heavy-fermion compounds and the high-$T_c$ cuprates, have an electrical resistivity that scales linearly with temperature, in stark contrast to a typical Fermi liquid. Such behavior appears to be intimately related to quantum criticality, \textit{i.e.} the presence of a continuous phase transition at zero temperature as a function of some non-thermal control parameter (pressure, magnetic field, etc.) \cite{sachdev,coleman,gegenwart}. The low-energy dynamics at a quantum critical point is generically described by some strongly-coupled, scale-invariant theory. That theory can also describe finite-temperature physics, as long as temperature is the largest scale, and hence can influence the finite-temperature resistivity.

Few reliable methods exist to compute physical observables for these systems, with real-time physics being particularly difficult to study. Lattice simulation is not reliable because either a finite chemical potential or Lorentzian signature, as required for real-time transport, generically introduces the ``sign problem,'' in which a complex phase renders Monte-Carlo sampling practically impossible.

An alternative method is the anti-de Sitter/Conformal Field Theory (AdS/CFT) correspondence, or more generally gauge-gravity duality \cite{Maldacena:1997re,Witten:1998qj,Gubser:1998bc}. Gauge-gravity duality is a holographic duality between a strongly-coupled non-Abelian gauge theory and a weakly-coupled theory of gravity in one higher dimension. The field theory ``lives'' on the boundary of the bulk holographic spacetime, for example if the bulk spacetime is AdS then the dual CFT lives on the AdS boundary. The field theory energy scale is dual to the radial/holographic direction. Every gauge-invariant field theory operator is dual to a field in the bulk. For example, the stress-energy tensor is dual to the metric, while the conserved current of a global $U(1)$ symmetry is dual to a bulk $U(1)$ gauge field. A thermal equilibrium state with finite entropy appears in the gravity theory as a black hole spacetime \cite{Witten:1998zw}. In the field theory, perturbations away from thermal equilibrium are described by a low-energy effective theory called hydrodynamics. Via gauge-gravity duality, the study of hydrodynamics becomes the study of small perturbations of black holes \cite{Bhattacharyya:2008jc}.

Gauge-gravity duality currently does not describe any real system, but is valuable nevertheless as one of the few ways to answer questions about finite-density, real-time physics in strongly-coupled systems. Moreover, gauge-gravity duality has revealed remarkable universality in the transport properties of such systems: any theory with a gravitational dual (with Einstein-Hilbert action, in a state with $SO(2)$ rotational symmetry \cite{Erdmenger:2010xm}) has the same, very small, ratio of shear viscosity $\eta$ to entropy density $s$, namely $\eta/s = 1/4\pi$ \cite{Kovtun:2004de}. That value is remarkably close to the value estimated for the QGP created at RHIC \cite{Luzum:2008cw}.

Holographic models fall into two categories, ``top-down'' and ``bottom-up.'' The former begin with a genuine string theory or supergravity action. In such cases the dual field theory can (usually) be determined precisely, allowing not only for specific translation between field theory and gravity, but also for comparison to weak-coupling results. The price is a large number of fields, often with complicated couplings. Bottom-up constructions begin with a simple bulk action, including few fields with simple couplings, to capture the essential physics. That provides technical simplicity without sacrificing universality. Whether a dual field theory exists or is well-defined is not guaranteed, however, so any results must be handled with care.

The simplest bottom-up holographic model of strongly-interacting matter is Einstein-Maxwell theory (with negative cosmological constant) \cite{Romans:1991nq}. The dual is some CFT with a global $U(1)$ symmetry. The AdS-Reissner-Nordstr\"om (AdS-RN) charged black hole solution describes thermal equilibrium states with a finite $U(1)$ charge density. The extremal limit of AdS-RN is dual to the field theory's zero-temperature limit at finite density, and is the paradigm of holographic quantum critical matter \cite{Hartnoll:2009sz,Hartnoll:2011fn}.

Extremal AdS-RN has a number of unusual properties. Chief among them is the presence of an extremal horizon, indicating that the entropy density does not vanish in the zero-temperature limit. In other words, the zero-temperature, finite-density state has a large degeneracy of microstates. That renders the system vulnerable to instabilities: generically, any small perturbation will break the degeneracy, and the system will settle into a new, presumably non-degenerate, ground state.

On the gravity side, the instability is intimately related to the presence of a $(1+1)$-dimensional AdS factor, $AdS_2$, in the near-horizon region \cite{Faulkner:2009wj}. A condition for the stability of a field in AdS is the Breitenlohner-Freedman (BF) bound \cite{Breitenlohner:1982jf}, a bound on how negative the mass (squared) of a bulk field can be. A bulk field may satisfy the BF bound in the near-boundary region but violate the $AdS_2$ BF bound. The exact nature of the instability depends on the bulk field content. For example, a charged scalar field may become unstable and develop a non-trivial profile: the horizon is replaced by ``scalar hair'' \cite{Hartnoll:2008kx}. The dual field theory state is then a zero-temperature superfluid in which the global $U(1)$ symmetry is spontaneously broken. The near-horizon $AdS_2$ factor indicates that in the field theory some (0+1)-dimensional CFT emerges at low energies.

Extremal AdS-RN also has unusual transport properties. Holographic calculations revealed that the dual theory has a finite shear viscosity, and that indeed $\eta/s=1/4\pi$ \cite{Edalati:2009bi}. The dual theory also admits a sound mode with speed $v$ identical to the value dictated by scale invariance, $v^2 = 1/d$ in $d$ spatial dimensions \cite{Edalati:2010pn}.

These results are mysterious. On general grounds we do not expect hydrodynamics to remain a reliable effective description at arbitrarily low temperatures. Hydrodynamics is an effective description in which we expand expectation values of conserved quantities, like the stress-energy tensor, in a small parameter, momentum times the mean free path. In the CFT the only meaningful parameter is the ratio of temperature to chemical potential, $T/\mu$. In the limit $T/\mu \gg 1$, we expect the mean free path to scale as $1/T$ plus corrections in $\mu/T$. As we decrease $T/\mu$, however, the mean free path grows, so we expect the hydrodynamic expansion to break down at some point. If we continue to use hydrodynamics beyond that point, we may see sick behavior. For example, if $\eta$ remains finite as $T/\mu \rightarrow 0$ then the divergence of the entropy current grows without bound \cite{Nickel:2010pr}. The holographic results for zero-temperature transport suggest that perhaps some effective theory, very much like hydrodynamics, exists in the limit $T/\mu \ll 1$. What theory replaces hydrodynamics at low temperatures? For an attempt to answer this question via a holographic version of the Wilsonian renormalization procedure, see ref.~\cite{Nickel:2010pr}.

Our goal is to perform a complete stability analysis in a top-down model of holographic matter at zero temperature. AdS-RN may be extracted from top-down holographic models (see for example refs.~\cite{Cvetic:1999xp,Buchel:2006gb,Gauntlett:2006ai,Gauntlett:2007ma} and references therein), which usually involve many fields with complicated couplings. To our knowledge no complete stability analysis, including all fields, has been performed in any top-down model.\footnote{Many top-down models rely on (consistent) truncations, in which the full set of supergravity fields is restricted to some subset of fields that only source fields within that subset. Obviously, only instabilities of fields within that subset can then be detected, that is, the truncation leaves us blind to instabilities of other fields, and so we cannot determine the true ground state with certainty.} We will therefore employ a tractable substitute: a probe D-brane.

We will study type IIB supergravity in the near-horizon geometry of very many D3-branes, $AdS_5 \times S^5$, with a single probe D7-brane extended along $AdS_5 \times S^3$ \cite{Karch:2002sh}. The dual field theory is $\N=4$ supersymmetric $SU(N_c)$ Yang-Mills (SYM) in the large-$N_c$ and large-coupling limits, coupled to a single $\N=2$ hypermultiplet in the fundamental representation of $SU(N_c)$, \textit{i.e.} flavor fields. The field theory has a global $U(1)$ flavor symmetry analogous to baryon number in QCD, which we will denote $U(1)_B$. We will introduce a baryon number chemical potential and an $\N=2$ supersymmetric mass for the hypermultiplet. The baryon number current and the flavor mass operator are dual to the $U(1)$ gauge field and a scalar field on the D7-brane worldvolume \cite{Kobayashi:2006sb}. The D7-brane action, including the Dirac-Born-Infeld (DBI) and Wess-Zumino (WZ) terms, describes the dynamics of these fields. Fortunately, exact solutions for these fields are known that describe field theory states with a finite baryon density at zero temperature, when the chemical potential is larger than the mass \cite{Karch:2007br}. We will study the spectrum of linearized fluctuations of worldvolume fields about these background solutions, dual to the spectrum of mesonic operators \cite{Kruczenski:2003be}.

Our system shares many features with extremal AdS-RN, in particular a finite entropy density and a sound mode with speed identical to the conformal value \cite{Karch:2008fa,Kulaxizi:2008kv,Karch:2009eb}. An $AdS_2$ region is also present in some sense, since fluctuations of worldvolume fields effectively ``see'' an $AdS_2$ region deep in the bulk, due to their couplings to the background worldvolume fields \cite{Jensen:2010ga,Jensen:2010vx, Evans:2010np,Nickel:2010pr}, as we will review in detail below.\footnote{If we leave the probe limit, and allow the D7-brane to back-react on the fields of supergravity, then we might expect to find an extremal black hole, which on general grounds we expect to have a finite entropy and near-horizon $AdS_2$ region. The features we see for the probe D7-brane may simply be the probe-limit remnants of those for the extremal black hole.}

We have four main reasons to expect instabilities in our system. The first is the finite entropy density at zero temperature. The second is the presence of charged scalars in the field theory susceptible to Bose-Einstein condensation. The third is the presence of charged fermions: our intuition from large-$N_c$ QCD with an asymptotically large chemical potential is that quarks may exhibit an instability towards an inhomogeneous ground state, the so-called ``chiral density wave'' \cite{Deryagin:1992rw,Shuster:1999tn}. The fourth reason comes from other holographic systems, where either bulk Chern-Simons or axion-like couplings triggered instabilities towards inhomogeneous ground states \cite{Nakamura:2009tf,Ooguri:2010kt,Ooguri:2010xs,Bayona:2011ab,Donos:2011bh,Bergman:2011rf}. The WZ terms in the D7-brane action indeed act as Chern-Simons couplings for the worldvolume gauge field.

Our analysis is complete in that we include fluctuations of \textit{all bosonic} D7-brane fields.\footnote{Fermionic fields also live on the D7-brane worldvolume \cite{Kirsch:2006he,Ammon:2010pg}, but on general grounds we do not expect them to become unstable, as explained for example in ref.~\cite{Faulkner:2009wj}, so we omit them from our analysis.} The only other fields in our system are those of supergravity, but in the probe limit these are insensitive to the D7-brane. We are thus including all relevant bosonic fields in our top-down model. Crucially, recall that none of the D7-brane worldvolume fields are charged under the worldvolume $U(1)$, or in field theory language mesons are not charged under baryon number, so we know \textit{a priori} that we will not see a superfluid instability. We include in our analysis fluctuations with finite momentum, in anticipation of instabilities towards inhomogeneous ground states. Our analysis is incomplete in that we do not study arbitrarily high momenta, although that is a limitation of our numerical procedure rather than a matter of principle.

Despite all the circumstantial evidence, we find no signs of any instabilities. This top-down model appears to be perturbatively stable, at least over a wide range of momenta. We hasten to add that this low-momentum stability is probably special to the large-$N_c$ and probe limits, as we discuss in what follows.

In our analysis we recover the ``zero sound'' mode of refs.~\cite{Karch:2008fa,Kulaxizi:2008kv}, appearing as a pole in the $U(1)_B$ charge density two-point function. We also discover a new mode with purely imaginary dispersion in the two-point function of a certain scalar operator that is a vector of the $SU(2)_R$ R-symmetry. We cannot resist drawing an analogy between this mode and the ``spin diffusion'' mode in some itinerant electron systems. Here we adopt the perspective of, for example, refs.~\cite{Iqbal:2010eh,Karch:2010mn}. In a system of electrons, at low energies spin-orbit interactions are suppressed and spin rotations decouple from spacetime rotations. Spin essentially becomes an internal $SU(2)$ quantum number, that is, a global symmetry, for which we may ask about the rate of diffusion. In our case, the $SU(2)$ is part of the R-symmetry, so we dub the mode we find ``R-spin diffusion.'' The presence of the zero sound and R-spin diffusion lend further evidence for the existence of an effective, hydrodynamic-like theory describing holographic quantum critical matter.

The absence of instabilities may be consistent with the interpretation, proposed in refs.\cite{Karch:2007br,Jensen:2010vd}, that the solution of ref.~\cite{Karch:2007br} (about which we perturb) represents scalars that have \textit{already} Bose-Einstein condensed. The argument proceeds as follows. Recall that an $\N=2$ hypermultiplet includes a Dirac fermion and two complex scalars, all with the same $U(1)_B$ charge. In our theory the scalars have quartic self-interactions as well as Yukawa and other couplings.\footnote{The Lagrangian of our theory is written explicitly in ref.~\cite{Chesler:2006gr}.} We then introduce a $U(1)_B$ chemical potential, which we begin to increase. When the chemical potential is less than the flavor mass, the ground state is indistinguishable from the vacuum: all operators have zero expectation value. Once the chemical potential equals the flavor mass, the scalars should condense, producing a nonzero density. Indeed, holographic calculations reveal a phase transition of precisely that character \cite{Karch:2007br}. In addition, the $\N=2$ supersymmetric completion of the hypermultiplet Dirac mass operator acquires an expectation value identical in form to that of condensed scalars with quartic self-interaction. The fact that we find no instabilities may be because the instability already occurred: the scalars condensed, and the system is already in the ground state. Such an interpretation has several problems, however, as we review in detail below. Chief among them is the fact that the scalars are charged under $SU(N_c) \times SU(2)_R \times U(1)_B$, so if they condense we expect these symmetries to be broken to some subgroup. That does not occur for the solution of ref.~\cite{Karch:2007br}. Our calculation serves to characterize further whatever state the solution of ref.~\cite{Karch:2007br} represents: the low-energy spectrum includes not only the zero sound mode but also the R-spin diffusion mode, and at very low energies a (0+1)-dimensional CFT emerges.

The time component of the extremal AdS-RN metric has a double zero. That generically introduces an irregular singular point in the equations of motion of fluctuations, making numerical solutions difficult to obtain. Two techniques designed to address this problem are matched asymptotic expansions, which is reliable for sufficiently low frequencies \cite{Faulkner:2009wj}, and Leaver's method \cite{Leaver:1990zz,Denef:2009yy}. In our system the fluctuation equations also involve irregular singular points. We employ matched asymptotic expansions, but Leaver's method is difficult to use, as we explain below. We thus develop a new numerical technique to compute the spectrum, which we call the ``zig-zag'' method (for reasons that will become clear), which may be applicable more generally.

This paper is organized as follows. In section~\ref{ss:theory} we briefly review the field theory and its holographic dual, the circumstantial evidence for instabilities, and the interpretation of the solution of ref.~\cite{Karch:2007br} as a condensate of scalars. In section~\ref{ss:fluctuations} we derive the equations of motion for fluctuations of D7-brane fields. In section~\ref{ss:spectrum} we perform matched asymptotic expansions, explain the zig-zag method, and present our numerical results for the spectrum. We conclude with some discussion in section~\ref{ss:conclusions}.

\section{The Theory and Its Holographic Dual}\label{ss:theory}

We study a strongly-coupled CFT, namely $\N=4$ SYM theory with gauge group $SU(N_c)$ in the large-$N_c$ limit and with large 't Hooft coupling $\lambda \equiv g_{YM}^2 N_c \gg 1$, where $g_{YM}$ is the SYM coupling. We introduce a single $\N=2$ supersymmetric hypermultiplet in the fundamental representation of the gauge group, which in analogy with QCD we will call a flavor field. The field content of an $\N=2$ hypermultiplet is a Dirac fermion $\psi$ and two complex scalars, $q$ and $\tilde{q}$. We will call the fermion a quark and the scalars squarks.

We work in the probe limit: with $N_f$ flavors of hypermultiplets, the probe limit consists of keeping $N_f$ fixed as $N_c \to \infty$, so that $N_f/N_c \ll 1$. If we work to leading order in that small parameter, then we may neglect quantum effects due to the flavor fields, such as the running of the coupling. For example, at weak coupling, $\lambda \ll 1$, the probe limit consists of discarding all diagrams with flavor fields in any loops.

$\N=4$ SYM theory has an $SO(6)_R$ R-symmetry. The $\N=2$ supersymmetric flavor fields have superpotential couplings that break the symmetry to $SO(4) \times U(1)_R \simeq SU(2)_L \times SU(2)_R \times U(1)_R$, of which the $SU(2)_R \times U(1)_R$ acts as the $\N=2$ R-symmetry. The $U(1)_R$ acts as a chiral symmetry for the quarks, while the squarks are neutral.

We will introduce an $\N=2$ supersymmetry-preserving mass $M$ for the flavor fields, which explicitly breaks the chiral $U(1)_R$, much like QCD. We will denote the $\N=2$ supersymmetric completion of the Dirac fermion mass operator as $\Om$. Of the three complex adjoint scalars of $\N=4$ SYM, only one is charged under $U(1)_R$. Denoting that scalar as $\Psi$, $\Om$ schematically takes the form (the precise operator is written in ref.~\cite{Kobayashi:2006sb})
\be
\label{eq:omdef}
\Om = i \bar{\psi} \psi + q^{\dagger} (M + \Psi) q + \tilde{q}^{\dagger} (M + \Psi) \tilde{q}.
\ee

The theory has a $U(1)_B$ flavor symmetry, analogous to baryon number in QCD. We will introduce a baryon number chemical potential $\mu$. Although we will mostly be interested in the theory at zero temperature, we will sometimes discuss the theory in a thermal equilibrium state with temperature $T$.

To obtain the holographic dual of the above theory we begin in type IIB string theory with the following supersymmetric intersection of $N_c$ D3-branes with a single D7-brane,
\be
\label{table:d3d7}
\begin{array}{c|cccccccccc}
   & x_0 & x_1 & x_2 & x_3 & x_4 & x_5 & x_6 & x_7 & x_8 & x_9\\ \hline
\mbox{D3} & \times & \times & \times & \times & & &  &  & & \\
\mbox{D7} & \times & \times & \times & \times & \times  & \times
& \times & \times &  &   \\
\end{array}
\ee
Open strings with both ends on the D3-branes give rise at low energies to $\N=4$ SYM, while open strings with one end on the D3-branes and one on the D7-branes give rise to the single $\N=2$ hypermultiplet. If we separate the D3- and D7-branes by a distance $L$ in the $(x_8,x_9)$ plane, then the open strings between them acquire a length and hence the flavor fields acquire a mass that is $L$ times the string tension, $M=L/(2\pi\ap)$. The relative angle in the $(x_8,x_9)$ plane represents the phase of the mass. The $U(1)_R$ symmetry corresponds simply to rotations in the $(x_8,x_9)$ plane.

Taking the usual Maldacena limits, we obtain the holographic dual of the large-$N_c$, large-coupling $\N=4$ SYM theory, type IIB supergravity in the near-horizon geometry of the D3-branes, $AdS_5 \times S^5$ with $N_c$ units of Ramond-Ramond (RR) five-form flux on the $S^5$. We will write the metric and five-form as
\ba\label{ads5}
ds^2 & = & G_{\mu \nu} dx^{\mu} dx^{\nu} = \frac{r^2}{R^2}(-dt^2+d\vec{x}^2)+\frac{R^2}{r^2}(dr^2+r^2d\Omega_5^2), \no
F^{(5)}& = &\frac{4}{R^4}(r^3dx^0\wg dx^1\wg dx^2\wg dx^3\wg dr)-4R^4d\Omega_5,
\ea
where $R$ is the $AdS_5$ radius of curvature, $r$ is the $AdS_5$ radial coordinate, with the boundary at $r \to\infty$, the field theory Minkowski coordinates are $t$ and $\vec{x}$, and $d\Omega_5$ is the volume form of $S^5$. We will use an $S^5$ metric of the form
\be
d\Omega_5^2=d\theta^2+\sin^2\theta d\phi^2+\cos^2\theta d\Omega_3^2,
\ee
where $d\Omega_3$ is the volume form of $S^3$. The RR four-form potential $C^{(4)}$ giving rise to the five-form via $F^{(5)} = dC^{(4)}$ then takes a simple form:
\be
C^{(4)}=\frac{1}{R^4}(r^4dx^0\wg dx^1\wg dx^2\wg dx^3+R^4\cos^4\theta d\phi\wg d\Omega_3).
\ee
The radius of curvature is fixed to be $R^4 = 4 \pi g_s N_c \alpha'^2$ where $g_s$ is the string coupling. In the AdS/CFT dictionary $4\pi g_s = g_{YM}^2$, so the large 't Hooft-coupling limit of the CFT corresponds to $R^4/\alpha'^2 \gg 1$, the limit in which excited string states become heavy relative to the $AdS_5$ scale.\footnote{Starting now, we will use units in which $R\equiv 1$. In these units, $\alpha' = \lambda^{-1/2}$ is dimensionless.} The AdS/CFT dictionary also equates the isometries of the bulk spacetime with the symmetries of the CFT: the isometry group of $AdS_5$ is dual to the (3+1)-dimensional conformal group, and the $SO(6)$ isometry of the $S^5$ is dual to the R-symmetry of $\N=4$ SYM.

In the near-horizon limit, the D7-brane is extended along $AdS_5 \times S^3$ inside $AdS_5 \times S^5$. The action for a D7-brane in this background includes the DBI and WZ terms,
\be\label{D7}
S_{D7}=-T_{D7}\int d^8\xi\sqrt{-\text{det}(g_{ab}+(2\pi\ap) F_{ab})}+\frac{(2\pi\ap)^2}{2}T_{D7}\int P[C^{(4)}]\wg F\wg F,
\ee
where $T_{D7} = \frac{g_s^{-1} \alpha'^{-4}}{(2\pi)^7}$ is the D7-brane tension, $(\xi^1,\ldots,\xi^8)$ are the worldvolume coordinates, $g_{ab}$ is the induced metric of the D7-brane, $F_{ab} = \partial_a A_b - \partial_b A_a$ is the field strength associated with the worldvolume $U(1)$ gauge field $A_a$, and $P[C^{(4)}]$ is the pullback of the RR four-form to the D7-brane worldvolume.

The statement that the D7-brane is dual to the $\N=2$ hypermultiplet means that every field on the D7-brane is dual to some mesonic operator, that is, a gauge-invariant operator bilinear in hypermultiplet fields. The fields living on the D7-brane are two scalars, representing fluctuations of the D7-brane in the two transverse directions, and the eight components of the $U(1)$ gauge field. These are dual to spin zero and spin one mesons. Fermionic fields also live on the D7-brane worldvolume, which are dual to fermionic mesons \cite{Kirsch:2006he,Ammon:2010pg}, but we will leave an analysis of their fluctuations for future work. Generically, higher-spin mesons are dual to heavy string excitations accessible only outside of the supergravity limit.

Consider for example $\Om$, which is dual to a D7-brane worldvolume scalar. Indeed, before the near-horizon limit that scalar represents the separation of the D3- and D7-branes in the overall transverse plane. To describe that here, we rewrite the metric of the $\mathbb{R}^6$ spanned by $r$ and the $S^5$ as
\be
dr^2 + r^2 d\Omega_5^2 = d\rho^2 + \rho^2 d\Omega_3^2 + dy^2 + y^2 d\phi^2,
\ee
where $\rho^2=r^2\cos^2\theta$ and $y^2 = r^2 \sin^2\theta$ and hence $r^2 = \rho^2 + y^2$. The D7-brane is extended along $\rho$ and the $S^3$. We have used polar coordinates $y$ and $\phi$ for the $(x_8,x_9)$ plane. On the D7-brane worldvolume $y$ and $\phi$ appear as scalar fields, which generically depend on all the D7-brane coordinates $(\xi^1,\ldots,\xi^8)$. The mode of $y$ that is independent of the $S^3$ coordinates is dual to $\Om$. To describe a constant real mass $M$, preserving space and time translation symmetries, we need a solution of the form $y(\rho)$. The solution to the equations of motion is then simply a constant, $y(\rho) = L$, which represents flavor fields with a mass $M=L/(2\pi\ap)$ \cite{Karch:2002sh,Kruczenski:2003be}.

In the trivial case, $y(\rho)=0$, the D7-brane clearly preserves the $SO(4) \times U(1)_R$ isometry of $S^3$ and $\phi$, which is dual to the symmetry preserved by a massless hypermultiplet. In the nontrivial case, $y(\rho)=L$, the D7-brane preserves only the $SO(4)$ isometry, dual to the symmetry preserved by a massive hypermultiplet.

The trivial solution describes a D7-brane that fills $AdS_5 \times S^3$. Massless flavors in the probe limit will indeed preserve conformal invariance, hence the appearance of the $AdS_5$ factor. Recalling that the $AdS_5$ radial coordinate is dual to the CFT energy scale, we can interpret the non-trivial solution $y(\rho)=L$ as the renormalization-group (RG) flow associated with massive flavors. Here the induced D7-brane metric is $AdS_5 \times S^3$ near the boundary $\rho \to \infty$, but as the D7-brane extends into $AdS_5$, the $S^3$ shrinks and eventually collapses to zero size (smoothly) at $\rho=0$. Recalling that the $AdS_5$ radial coordinate is not $\rho$ but $r = \sqrt{\rho^2 + y(\rho)^2}$, we see that at $\rho=0$ the D7-brane has only reached $r = L$. From the perspective of an observer in $AdS_5$, the D7-brane simply ends there. In the RG flow, the flavor fields are absent at energy scales below their mass gap.

The spectrum of fluctuations about the non-trivial solution $y(\rho)=L$ was computed in ref.~\cite{Kruczenski:2003be}. In that case the fluctuations are normal modes, \textit{i.e.} standing waves trapped between the $AdS_5$ boundary and the endpoint of the D7-brane. These normal modes are dual to poles in the two-point functions of the dual operators, \textit{i.e.} mesons, which in the large-$N_c$ and probe limits are exactly stable. These mesons do not arise due to chiral symmetry breaking or confinement, rather they are simply deeply bound states, with masses on the order of $M/\sqrt{\lambda}$ \cite{Kruczenski:2003be}. We will not review the meson spectrum in detail here, but we will mention that the mesons form representations of the $SO(4)$ symmetry, and in fact because the mesons form short multiplets of the $\N=2$ superconformal algebra, their dimensions are completely determined by their R-symmetry charges \cite{Kruczenski:2003be}. The dual statement is that fluctuations of D7-brane fields can be decomposed into spherical harmonics of the $S^3$, which cannot mix since they form different representations of $SO(4)$, and higher $S^3$ angular momentum implies larger (Kaluza-Klein) mass, which means the dual operator has a larger dimension.

\subsection{Finite Density States}
\label{ss:finitedensitystates}

The conserved current associated with baryon number $U(1)_B$, $J^{\nu}$, is dual to the components of the worldvolume gauge field in the Minkowski directions that are independent of the $S^3$ coordinates, $A_{\nu}$ \cite{Kobayashi:2006sb}. In particular, if we want to study states with a finite chemical potential $\mu$ and charge density $\jt$, preserving space and time translations, we must introduce a worldvolume $A_t(\rho)$. We can also motivate the introduction of $A_t(\rho)$ with the following physical argument. A single charged excitation is represented in the bulk by a single string that ends on the D7-brane. To obtain a charge density, we must introduce a density of strings, which will source the worldvolume gauge field, producing an electric field in the radial direction, $F_{\rho t}(\rho)$. If we choose $A_{\rho}=0$ gauge, then to obtain $F_{\rho t}(\rho)$ we must introduce $A_t(\rho)$.

Notice that a number on the order of $N_c$ strings is strong enough to pull the D7-branes all the way to the ``bottom'' of $AdS_5$, meaning $r=0$, hence any solution representing a finite baryon density of order\footnote{Following ref.~\cite{Karch:2007br}, we employ a normalization in which a single charge/string has baryon number one, so that a baryon actually has baryon number $N_c$.} $N_c$ will not end, \textit{i.e.} the $S^3$ will not collapse at finite $r$ \cite{Kobayashi:2006sb}. In such cases the fluctuations of worldvolume fields will not be normal modes but quasi-normal modes (QNM's), that is, the eigenfrequencies will become complex, since some part of the standing wave can now ``leak'' across the (Poincar\'e) horizon of $AdS_5$. That conforms to our field theory expectation that any finite density will give mesons a finite width.

Luckily, two exact solutions describing finite $M$ and $\mu$ are known \cite{Karch:2007br}. The first is simply $A_t(\rho)=\mu$ with constant $y(\rho) = L$. These constant fields are a solution of the equation of motion because the field strength $F_{\rho t}=0$, so the gauge field equation is trivially satisfied and the scalar equation is unchanged from the $A_t(\rho)=0$ case. These solutions describe zero density states, however, and so are only relevant when studying the grand canonical ensemble. The second solution can be written in terms of an incomplete Beta function,
\ba\label{KO}
y(\rho) & = & \frac{1}{6}c{\cal N}^{-1/3}\(\frac{d^2}{(2\pi\ap)^2}-c^2\)^{-1/3} B\left(\frac{{\cal N}^2\rho^6}{{\cal N}^2\rho^6+\frac{d^2}{(2\pi\ap)^2}-c^2};\frac{1}{6},\frac{1}{3}\right), \no
A_t(\rho) & = & \frac{1}{(2 \pi \alpha')}\frac{1}{\veps}y(\rho),
\ea
where $\N=T_{D7}2\pi^2=\frac{\lambda N_c}{(2\pi)^4}$, and the parameters $c$ and $d$ are fixed to be
\be
\label{eq:cddefs}
&&c = \gamma{\cal N}(2\pi\alpha')^3\left(\mu^2-M^2\right)M, \\
&&d = \gamma{\cal N}(2\pi\alpha')^4\left(\mu^2-M^2\right)\mu \nonumber,
\ee
with $\gamma=\left[\frac{1}{6}B\(\frac{1}{6},\frac{1}{3}\)\right]^{-3}\approx0.363$. These are related to the expectation values of the dual operators via $\Omv = (2\pi\ap)c$ and $\jt=d$. We have also introduced the parameter
\be
\label{eq:veps}
\veps = \frac{c(2\pi\ap)}{d} = \frac{\Omv}{\jt} = \frac{M}{\mu}.
\ee
Notice that $y(\rho=0)=0$, indicating that the $S^3$ collapse occurs only when $r=\sqrt{\rho^2 + y(\rho)^2}$ is zero, that is, at the bottom of $AdS_5$. Notice also that although $M=0$ implies $c=0$ and hence $y(\rho)=0$, the gauge field $A_t(\rho) = \frac{1}{\veps} y(\rho)$ remains nontrivial.

For equilibrium states, the AdS/CFT dictionary equates the on-shell bulk action with the field theory grand canonical partition function. By comparing the free energies associated with the two solutions above, we discover a phase transition \cite{Karch:2007br}. When $\mu<M$, the thermodynamically preferred solution is the trivial one, describing states with $\Omv=0$ and zero density, $\jt=0$. When $\mu\geq M$, the preferred solution is the non-trivial one, describing states with the nonzero $\Omv$ and $\jt$ given by the $c$ and $d$ in eq.~\eqref{eq:cddefs}. The critical exponents associated with the transition take mean-field values. Technically, this zero-temperature transition is a quantum phase transition, albeit a rather boring one from nothing (zero density) to something (finite density). Notice that the transition is only apparent in the grand canonical ensemble. In the canonical ensemble we have access only to finite-density states, by definition. Notice also that the parameter $\veps$ defined in eq.~\eqref{eq:veps} is bounded: $0 < \veps < 1$. The limit $\veps = 0$ corresponds to zero mass, or finite mass and infinite chemical potential. The limit $\veps = 1$ corresponds to $M = \mu$, where the transition to zero-density states occurs.

Upon introducing a finite temperature, dual holographically to studying the same system but in an AdS-Schwarzschild black hole spacetime, straightforward calculations reveal two unusual features at temperatures low compared to the density. First, the heat capacity scales with temperature as $c_V \propto T^6$, in stark contrast to a Fermi liquid, where $c_V \propto T$, or a Bose liquid, where $c_V \propto T^3$ in three spatial dimensions \cite{Karch:2008fa}. Second, the zero-temperature limit of the entropy density $s$ is nonzero: $\lim_{T \to 0} s = \frac{1}{2}\sqrt{\lambda} \jt$ \cite{Karch:2008fa}. As argued in ref.~\cite{Karch:2009eb}, these are essentially the heat capacity and entropy associated with a single charge/string, times the number of charges/strings, $\jt$. The low-temperature, finite-density states appear to be locally thermodynamically stable \cite{Benincasa:2009be}.

To our knowledge, the only analyses of the fluctuation spectrum about the zero-temperature, finite-density solution appear in refs.~\cite{Karch:2008fa,Kulaxizi:2008kv,Kim:2008bv},\footnote{With finite baryon density and finite temperature, various parts of the fluctuation spectrum have been computed in refs.~\cite{Erdmenger:2007ja,Myers:2008cj,Mas:2008jz,Erdmenger:2008yj,Mas:2008qs,Mas:2009wf,Kaminski:2009ce,Kaminski:2009dh,Shock:2009fr}.} where a pole was discovered in the two-point function of the $U(1)_B$ charge density (and current) two-point functions, which for massless flavors $M=0$ has a dispersion relation\footnote{Our frequency and momenta, $\omg$ and $k$, have signs opposite to those of ref.~\cite{Karch:2008fa}. In our conventions, the zero sound appears in the \textit{upper}-half of the complex $\omega$ plane, hence the imaginary part has a sign opposite to that of ref.~\cite{Karch:2008fa}.}
\be
\label{eq:zsdispersion}
\omega(k) = \pm \frac{1}{\sqrt{3}} k + i \frac{k^2}{6 \mu} + O\left( k^3\right),
\ee
which is formally identical to that of Landau's ``zero sound'' pole in the density two-point function in a Fermi liquid, hence the mode was dubbed zero sound by analogy. Notice in particular that the speed is identical to the speed of sound in a finite-temperature conformal plasma in (3+1) dimensions.

Our objective is to analyze fluctuations of all bosonic worldvolume fields, including finite-momentum fluctuations, in order to detect perturbative instabilities. Before proceeding to the details of the calculation, let us review the circumstantial evidence in favor of instabilities.

\subsection{Reasons to Expect Instabilities}

We have a number of reasons to expect instabilities, based on intuition from similar theories at weak coupling and also from similar holographic systems. 

The first reason is simply the presence of a nonzero entropy density at zero temperature, which renders the system vulnerable to instabilities.

The second reason is the presence of charged fermions in our field theory. Our intuition here comes from QCD with an asymptotically large baryon chemical potential. In such a regime we expect Fermi surfaces of weakly-interacting quarks (technically Landau quasi-particles). When $N_c=3$, standard arguments from the BCS theory of superconductivity suggest that the quarks form Cooper pairs, producing a color superconductor.\footnote{For a review of the general arguments, see ref.~\cite{Rajagopal:2000wf} and references therein.} In the large-$N_c$ limit, however, things may change. The quarks still interact weakly, but the diagram describing the relevant attractive interaction between them, providing the ``glue'' for the Cooper pairs, is non-planar, and the superconducting gap goes as $e^{-\sqrt{N_c}/\lambda}$, so both are suppressed in the 't Hooft limit \cite{Deryagin:1992rw,Shuster:1999tn}. One proposal for the actual ground state \cite{Deryagin:1992rw} is for quarks to pair with holes in the Fermi surface, moving in parallel with momenta on the order of $\mu$, producing a gauge-invariant condensate, the usual chiral condensate, but spatially modulated with a wave vector of magnitude $2 \mu$. In our theory, the $\N=2$ hypermultiplet includes fermions charged under $U(1)_B$ which may be susceptible to such an instability. Whether our finite-density states include a Fermi surface is unclear, however: the system has a zero sound mode, like a Fermi liquid, but a heat capacity unlike a Fermi or Bose liquid.

Inhomogenous ground states have also been found in several models of holographic matter, including probe brane systems \cite{Nakamura:2009tf,Ooguri:2010kt,Ooguri:2010xs,Bayona:2011ab,Bergman:2011rf,Donos:2011bh}. In these cases, the crucial ingredient was a Chern-Simons term in a (4+1)-dimensional bulk spacetime \cite{Nakamura:2009tf,Ooguri:2010kt,Ooguri:2010xs,Bayona:2011ab} or an axion-like coupling in a (3+1)-dimensional bulk spacetime \cite{Bergman:2011rf,Donos:2011bh}. Our model includes Chern-Simons-like terms via the D7-brane WZ terms, providing yet another suggestion that our system may have a finite-momentum instability. We should mention, however, that in our analysis below the WZ terms only enter the equations of motion for the D7-brane gauge field components in the $S^3$ directions, not in the $AdS_5$ directions (see eq.~\eqref{eome4}).

Another major reason to expect instabilities is the presence of charged scalars in our field theory, for which a chemical potential generically triggers Bose-Einstein condensation. The canonical example of Bose-Einstein condensation is a massive complex scalar, which necessarily has a global $U(1)$ symmetry, with quartic self-interactions. A chemical potential for the $U(1)$ effectively acts as a negative mass-squared. A chemical potential larger than the mass triggers a second-order phase transition: the potential develops a new minimum, the scalar acquires an expectation value, and the $U(1)$ symmetry is broken.

In our field theory, the hypermultiplet contains two complex scalars, the squarks, which indeed have quartic self-interactions, as well as Yukawa and other interactions \cite{Chesler:2006gr}. The two squarks $q$ and $\tilde{q}^{\dagger}$ are in the $N_c$ of $SU(N_c)$, form a doublet of $SU(2)_R$, carry identical $U(1)_B$ charges, and are neutral under $SU(2)_L \times U(1)_R$. If the squarks acquire an expectation value, then we expect $SU(N_c) \times SU(2)_R \times U(1)_B$ to be broken to some subgroup.

What would squark condensation look like holographically? Consider the intersecting branes in table~\eqref{table:d3d7}. Upon taking the usual Maldacena limits, we obtain $AdS_5 \times S^5$, where we may imagine that the D3-branes are ``hidden'' behind the $AdS_5$ Poincar\'e horizon. We then introduce $N_f$ D7-branes with worldvolume $U(1)$ electric flux. The D3-branes feel two forces in the radial direction, one towards the horizon, due to gravity (AdS is a potential well), and one towards the boundary, due to the electric flux. If the electric flux is sufficiently large, then the D7-branes may ``suck'' D3-branes out of the stack of $N_c$. The authors of ref.~\cite{Hartnoll:2009ns} called this ``Fermi seasickness.''

In our case, the D3-branes may then dissolve into the D7-branes, becoming instantons. The instanton moduli, namely the position, orientation in $U(N_f)$, and size, are isomorphic to the scalar expectation values, including those of the squarks and the adjoint scalars of $\N=4$ SYM \cite{Witten:1995gx,Douglas:1995bn,Erdmenger:2005bj,Apreda:2005yz}. Specifically, the size modulus is isomorphic to the squark expectation value. Normally, for a single supersymmetric D7-brane, the $U(1)$ worldvolume theory admits only point-like instantons, that is, instantons with zero size modulus, dual to squarks with zero expectation value. In the presence of nonzero background electric flux, however, the instanton could possibly acquire a finite size.\footnote{The possibility of finite-size instantons when $N_f=1$ should generically occur whenever the two stacks of branes feel a force towards one another. For example, a nonzero Kalb-Ramond B-field, which introduces non-commutativity on the worldvolume of the D7-branes, will also attract the D3-branes to the D7-branes, and allows for finite-size instantons even when $N_f=1$. In the holographic context, the B-field is dual to a Fayet-Iliopoulos term in the field theory \cite{Ammon:2008va}. We thank D.~Tong for discussions on this point.} The instanton number counts the number of D3-branes that have dissolved, and hence indicates breaking of $SU(N_c)$: a single-instanton solution implies the breaking $SU(N_c) \to SU(N_c-1) \times U(1)$. We thus expect a squark condensation instability to appear in the bulk as the D7-brane sucking D3-branes out of the stack, which then dissolve into the D7-brane, becoming finite-size instantons. Crucially, however, the \textit{rate} at which the D7-branes suck D3-branes out of the stack goes to zero when $N_c \rightarrow \infty$ and $N_f \ll N_c$ \cite{Hartnoll:2009ns}, that is, the process takes forever. We will thus not be able to detect this instability, since we work in the probe limit.

What symmetry-breaking instabilities can we detect, in principle, from our holographic calculation? We focus only on fluctuations of D7-brane worldvolume fields, dual to mesonic operators. These are invariant under $SU(N_c)$ and $U(1)_B$, so we cannot detect breaking of these symmetries. The various mesons of our theory form various representations of the remaining global $SU(2)_L \times SU(2)_R$ symmetry, and the $U(1)_R$ symmetry present when $M=0$, so we may detect instabilities toward states in which these symmetries are broken to some subgroups. We may also detect instabilities towards states breaking translational or rotational symmetries, for example if we see an instability towards condensation of vector mesons, or a finite-momentum instability.

In AdS/CFT we only have access to gauge-invariant observables. If we do see an instability, then most likely we will not be able to identify the microscopic mechanism. For example, in principle we could detect an instability towards a state in which $\Omv$ has a plane-wave form with wave vector of magnitude $2 \mu$, just like the chiral density wave in large-$N_c$ QCD. We would probably not be able to determine whether the microscopic origin of the instability in the field theory was the same as in large-$N_c$ QCD, that is, whether quarks were pairing with holes in a Fermi surface. AdS/CFT provides access only to ``macroscopic'' data, namely correlation functions of gauge-invariant operators.

Nevertheless, a microscopic interpretation of the solution in eq.~\eqref{KO} was proposed in refs.~\cite{Karch:2007br,Jensen:2010vd}. The suggestion is that the solution in eq.~\eqref{KO} in fact represents squarks that have \textit{already} condensed. The heuristic field theory argument goes as follows. We begin with nonzero $M$ but $\mu=0$, where the density $\jt = 0$ and also $\Omv = 0$. We then increase $\mu$ until $\mu > M$, at which point the scalars condense, producing a nonzero density $\jt$ and nonzero $\Omv$. Indeed, the form of $\Omv$ in eq.~\eqref{eq:cddefs} is identical to that of a charged scalar with quartic coupling. Moreover, we may anticipate that the system is perturbatively stable because, in some sense, the instability already took place: the squarks Bose-Einstein condensed.

A number of questions arise about such an interpretation, however. As mentioned above, if we trust our intuition from the canonical example of Bose-Einstein condensation, then if the squarks condense we expect $SU(N_c) \times SU(2)_R \times U(1)_B$ to be broken to some subgroup. The solution in eq.~\eqref{KO} describes a state in which the only operators that acquire expectation values, $J^t$ and $\Om$, are neutral under these symmetries. The dispersion relation in eq.~\eqref{eq:zsdispersion} looks suspiciously like that of a Goldstone boson, associated with $U(1)_B$ breaking for example, but as argued in ref.~\cite{Nickel:2010pr}, such a Goldstone boson would have a damping coefficient suppressed by a factor of $N_c$ relative to that of eq.~\eqref{eq:zsdispersion}. In short, the solution in eq.~\eqref{KO} preserves all global symmetries. The solution of eq.~\eqref{KO} also has instanton number zero, that is, the D7-brane has zero D3-brane charge, and hence no instanton size modulus. That suggests the squarks have zero expectation value and that $SU(N_c)$ is not broken. How can such a solution represent squark condensation?

Of course, our intuition may fail, due to an essential difference between our theory and the canonical example of Bose-Einstein condensation: the squarks are not gauge-invariant, so if they develop an expectation value, identifying the broken symmetries requires some caution. The $SU(N_c)$ gauge symmetry of our theory is not a physical symmetry at all, but merely a redundancy of our description, so no physical, gauge-invariant order parameter exists that can detect its breaking. In a weakly-coupled theory, such as the electroweak theory, a scalar expectation value merely serves as a ``convenient fiction'' to identify the breaking (to quote ref.~\cite{Rajagopal:2000wf}). A physical statement of spontaneous breaking of gauge invariance is, for example, the presence in the spectrum of a longitudinal mode for a vector boson. To detect breaking of a global symmetry, we must find a gauge-invariant charged operator and demonstrate that its expectation value is nonzero. For example, we can use the scalar expectation value itself to construct a gauge-invariant charged operator, by contracting its gauge indices with copies of itself \cite{Rajagopal:2000wf}.

We may thus be able to argue that, in fact, the squarks can condense without breaking any symmetries. Consider for example the $U(1)_B $ symmetry. In our theory the only gauge-invariant operator charged under $U(1)_B$ is the baryon operator itself,\footnote{Actually, the baryon is the only \textit{local} operator charged under $U(1)_B$. Non-local operators charged under $U(1)_B$ should also exist. We thank A.~Cherman for discussions on this point.} which due to anti-symmetry in color indices is actually constructed from $N_c$ quarks, hence we cannot build a baryon from a squark expectation value.\footnote{If $N_f \geq N_c$, we can construct a baryon using the squarks by anti-symmetrizing in both color and flavor indices. We have $N_f = 1 \ll N_c$, however.} We thus learn a very important lesson: in our theory, a nonzero squark expectation value does not immediately imply that $U(1)_B$ is broken, a fact that conflicts with our intuition from the canonical example. To be clear, our point is not that $U(1)_B$ can never break, but rather that a nonzero squark expectation value does not provide sufficient information to demonstrate $U(1)_B$ breaking.\footnote{A calculation of the back-reaction of D7-branes, with $U(1)$ electric flux, on the fields of supergravity appears in ref.~\cite{Chen:2009kx}. Although apparently the ansatz used in that calculation was incomplete, as shown in ref.~\cite{Bigazzi:2011it}, the results suggested that the electric flux produces a potential for the size modulus of D7-brane instantons, in keeping with the field theory expectation that the $U(1)_B$ density should produce a nonzero squark expectation value. The authors of ref.~\cite{Chen:2009kx} argued, however, that $U(1)_B$ is not broken when the squarks acquire an expectation value, which is consistent with our arguments.}

Even if we could somehow argue that the squarks condense without breaking any symmetries, we should ask how a system of condensed scalars can have a zero sound excitation and a heat capacity $c_V \propto T^6$. Both of these features seem strange, if we trust our intuition for weakly-coupled scalars. Of course, again our intuition may fail: the zero sound and unusual heat capacity may be strong-coupling effects.

Determining what microscopic state the solution in eq.~\eqref{KO} actually represents is beyond the scope of this paper, but we can further characterize that state by asking about the spectrum of fluctuations. We now proceed to the calculation of that spectrum.

\section{Fluctuations of D7-brane Fields}\label{ss:fluctuations}

We want to calculate the spectrum of fluctuations of bosonic D7-brane worldvolume fields. To do so we must determine the equations of motion for the fluctuations about the background solution described in section~\ref{ss:finitedensitystates}. Our analysis will closely follow that of ref.~\cite{Kruczenski:2003be}, where the spectrum of fluctuations about the constant solution $y(\rho)=L$ with $A_t(\rho)=0$ was computed, corresponding to the meson spectrum for finite-mass flavors in states with zero density.

We will employ a very general ansatz for the fluctuations, allowing them to depend on the Minkowski directions as well as the radial direction and the angles on the $S^3$. Specifically, we introduce a fluctuation of the embedding, $y(\rho) \rightarrow y(\rho) +(2\pi\ap) \chi$, of the other worldvolume scalar, $\phi=(2\pi\alpha') \varphi$, and of the worldvolume gauge field, $A_b=\dlt_b^tA_t+\dlt A_b$. The corresponding fluctuations of the induced metric and the $U(1)$ gauge field, 
up to quadratic order, take the following form:
\begin{eqnarray}\label{var_DBI}
\dlt g_{ab}&=&\dlt\left(\frac{\del x^\mu}{\del\xi^a}\frac{\del x^\nu}{\del\xi^b}\right)G_{\mu\nu} \no
&=&(2\pi\ap)\frac{y'}{r^2}(\del_a\chi\dlt_b^\rho+\del_b\chi\dlt_a^\rho)+(2\pi\ap)^2 \frac{1}{r^2}
(\del_a\chi\del_b\chi+\del_a\varphi\del_b\varphi), \no
\dlt F_{ab}&=&\del_a\dlt A_b-\del_b\dlt A_a.
\end{eqnarray}
We will only need the the pullback of the four-form to first order in the fluctuations,
\be\label{var_C4}
\dlt P[C^{(4)}] = \cos^4\theta(2\pi\ap)\(\del_\rho\(\frac{\varphi}{y}\)d\rho\wg d\Omega_3+\frac{1}{y}\del_t\varphi dt \wg d\Omega_3+\frac{1}{y}\vec{\nabla}\varphi d\vx\wg d\Omega_3\).
\ee

Inserting eqs.~\eqref{var_DBI} and \eqref{var_C4} into the D7-brane action eq.~\eqref{D7}, we find that the terms linear in fluctuations vanish because the background fields are on-shell, as expected. To write the terms quadratic in fluctuations, $S_{D7}^{{\cal O}(\delta^2)}$, let us first define, in terms of the background solution,
\be\label{etadndn}
\eta_{ab}\equiv g_{ab} + (2\pi \alpha') F_{ab} = 
\begin{pmatrix}
-r^2& -\frac{1}{\veps}y'& & \\
\frac{1}{\veps}y'& \frac{1}{r^2}(1+y'^2)& & \\
& & r^2 \delta_{\alpha \beta}& \\
& & & \frac{\rho^2}{r^2}\tilde{g}_{ij}
\end{pmatrix},
\ee
where $\a,\b,\g$ denote Minkowski directions, $i,j,k$ denote $S^3$ directions, with $\tilde{g}_{ij}$ the metric of the $S^3$, and we have ordered the coordinates as $(t,\rho,\a,i)$. The inverse of $\eta_{ab}$ will be denoted $\eta^{ab}$, for which we will define symmetric and anti-symmetric parts as
\be
\eta^{ab}_S \equiv \frac{1}{2} \left(\eta^{ab} + \eta^{ba} \right), \qquad \eta_A^{ab} = \frac{1}{2} \left(\eta^{ab} - \eta^{ba} \right).
\ee
Notice that the only non-vanishing element of $\eta^{ab}_A$ is $\eta^{\rho t}_A = - \eta^{t \rho}_A$. The terms in the action quadratic in fluctuations are then
\be
\label{L2}
&&S_{D7}^{{\cal O}(\delta^2)} \propto -\sqrt{-\eta}\frac{1}{2}\eta^{ba}\frac{1}{r^2}(\del_a\chi\del_b\chi+\del_a\varphi\del_b\varphi)-\sqrt{-\eta}\left (\frac{1}{8}\eta^{ba}\eta^{dc}-\frac{1}{4}\eta^{da}\eta^{bc} \right)\times \no
&&\left [\frac{y'}{r^2}(\del_a\chi\dlt_b^\rho+\del_b\chi\dlt_a^\rho)+\del_a\dlt A_b-\del_b\dlt A_a \right ] \left [\frac{y'}{r^2}(\del_c\chi\dlt_d^\rho+\del_d\chi\dlt_c^\rho)+\del_c\dlt A_d-\del_d\dlt A_c \right] \no
&&+r^4(\del_\rho\dlt A_i-\del_i\dlt A_\rho)\del_j\dlt A_k\eps^{ijk}-\frac{(2\pi\ap)}{\veps}\cos^4\theta\frac{y'}{y}\del_{\a}\varphi(\del_{\b}\dlt A_{\g}-\del_{\g}\dlt A_{\b})\eps^{\a\b\g},
\ee
where we have dropped overall constant factors, including for example a factor of $(2\pi\ap)^2 T_{D7}$, since these will not affect the equations of motion. Varying eq.~(\ref{L2}), and then making the gauge choice $\dlt A_\rho=0$, we find the equations of motion for $\varphi$, $\chi$, $A_{\rho}$, $A_t$, $A_{\a}$, and $A_i$, respectively:
\begin{subequations}
\be\label{EOM_ex}
&&\del_a(\sqrt{-\eta}\eta^{ab}_S\frac{1}{\rho^2+y^2}\del_b\varphi)=0,
\ee
\be
\label{eome1}
&&\del_a(\sqrt{-\eta}\eta^{ab}_S\frac{1}{\rho^2+y^2}\del_b\chi)-\sqrt{-\eta}(\eta^{\rho t})^2\frac{y'{}^2}{(\rho^2+y^2)^2}\del_t^2\chi-\del_a(\sqrt{-\eta}\eta^{\rho\rho}\eta^{ab}_S\frac{y'{}^2}{(\rho^2+y^2)^2}\del_b\chi) \no
&&+\del_a(\sqrt{-\eta}\eta^{\rho t}\eta^{ab}_S\frac{y'}{\rho^2+y^2}\del_b\dlt A_t)-\del_a(\sqrt{-\eta}\eta^{\rho t}\eta^{ab}_S\frac{y'}{\rho^2+y^2}\del_t\dlt A_b)=0,
\ee
\be
\label{eomc}
&&\sqrt{-\eta}\eta^{\rho\rho}\eta^{\rho t}\frac{y'}{\rho^2+y^2}\del_\rho\del_t\chi+\del_a(\sqrt{-\eta}\eta^{\rho\rho}\eta^{ab}_S\del_\rho\dlt A_b)=0,
\ee
\be
\label{eome2}
&&-\sqrt{-\eta}\eta^{\rho t}\eta^{tt}\frac{y'}{\rho^2+y^2}\del_t^2\chi+\del_a(\sqrt{-\eta}\eta^{\rho t}\eta^{ab}_S\frac{y'}{\rho^2+y^2}\del_b\chi)+\del_a(\sqrt{-\eta}\eta^{tt}\eta^{ab}_S\del_b\dlt A_t) \no
&&-\del_a(\sqrt{-\eta}\eta^{tt}\eta^{ab}_S\del_t\dlt A_b)=0,
\ee
\be
\label{eome3}
&&-\sqrt{-\eta}\eta^{\rho t}\eta^{xx}\frac{y'}{\rho^2+y^2}\del_t\vec{\nabla}\chi-\del_a(\sqrt{-\eta}\eta^{xx}\eta^{ab}_S\vec{\nabla}\dlt A_b)+\del_a(\sqrt{-\eta}\eta^{xx}\eta^{ab}_S\del_b\dlt\vec{A})=0
\ee
\be
\label{eome4}
&&-\sqrt{-\eta}\eta^{\rho t}\eta^{ij}\frac{y'}{\rho^2+y^2}\del_t\del_j\chi-\del_a(\sqrt{-\eta}\eta^{ab}_S\eta^{ij}\del_j\dlt A_b)+\del_a(\sqrt{-\eta}\eta^{ij}\eta^{ab}_S\del_b\dlt A_j) \no
&&-4(\rho^2+y^2)(\rho+yy')\eps^{ijk}\del_j\dlt A_k=0.
\ee
\end{subequations}
With our choice of gauge, the equation of motion for $\delta A_{\rho}$, eq.~\eqref{eomc}, becomes a constraint on the remaining fields. Notice that the final term in eq.~\eqref{L2} does not contribute to the equations of motion, due to the anti-symmetry of $\epsilon^{\a\b\g}$. In other words, the WZ terms only affect the $S^3$ gauge field fluctuations $A_i$. That is in contrast to ref.~\cite{Nakamura:2009tf}, for example, where the Chern-Simons terms entered the $AdS_5$ gauge field equations of motion. Notice also that if we set $F_{ab}=0$ in our $\eta_{ab}$ and also set $y'(\rho)=0$, then our equations reduce to those of ref.~\cite{Kruczenski:2003be}, up to a difference in gauge: we use $\delta A_{\rho}=0$ gauge, while in ref.~\cite{Kruczenski:2003be} the gauge choice $-\partial_t \delta A_t + \partial_{\a} \delta A_{\a} = 0$ was used.

Given that the background solution preserves translational and rotational invariance in the Minkowski directions, we will assume without loss of generality that the fluctuations have a plane-wave form with momentum only in the $x^3$ direction\footnote{Notice our choice of signs: we use $e^{i\omg t-ikx_3}$ rather than $e^{-i\omg t+ikx_3}$. With our conventions, stable modes will appear in the \textit{upper}-half of the complex $\omg$ plane.}, $e^{i\omg t-ikx_3}$. In that case, the transverse gauge field fluctuations $\dlt A_1$ and $\delta A_2$ decouple from the other fluctuations, and in fact obey exactly the same equation as $\varphi$, eq.~\eqref{EOM_ex}. The remaining fluctuations remained coupled to one another.

Following ref.~\cite{Kruczenski:2003be}, we will decompose the fluctuations into $S^3$ spherical harmonics and look for ans\"atze that decouple fields from one another. Let $\Y^m$ and $\Y^{m,\pm}_i$ be scalar and vector spherical harmonics on $S^3$, which obey the following relations:
\be\label{S3har}
\nabla^i\nabla_i\Y^m&=&-m(m+2)\Y^m, \\
\nabla^i\nabla_i \Y^{m,\pm}-R^k_{~j}\Y^{m,\pm}&=&-(m+1)^2\Y^{m,\pm}_j, \nonumber \\
\eps^{ijk}\nabla_j\Y^{m,\pm}_k&=&\pm(m+1)\sqrt{\tilde g}\tilde{g}^{ij}\Y^{m,\pm}_j, \nonumber \\
\nabla^i\Y^{m,\pm}_i&=&0, \nonumber
\ee
where $\nabla^i$, $R^i_{~j}=2 \delta^i_{~j}$, and $\tilde{g}$ are the Laplacian, Ricci tensor, and determinant of the $S^3$ metric. With respect to the $SO(4)= SU(2)_R \times SU(2)_L$ isometry of the $S^3$, the scalar harmonics $\Y^m$ transform in the $(\frac{m}{2},\frac{m}{2})$ representation, with $m \geq 0$. We can construct a vector harmonic from the scalar harmonic by taking $\nabla_i \Y^m$. The vector harmonics $\Y^{m,\pm}_i$ transform in the $(\frac{m\mp 1}{2},\frac{m \pm 1}{2})$ representation, with $m \geq 1$. We can decompose the scalars $\varphi$ and  $\chi$, and the gauge field components $A_{\a}$, into $S^3$ scalar spherical harmonics, and the gauge field components $A_i$ into vector spherical harmonics.

The ans\" atze that ultimately decouple the fluctuations are
\begin{subequations}\label{class}
\begin{align}
\label{eq:ansatz1}
&\chi=\Phi(\rho)e^{i\omg t-ikx_3}\Y^m,\,\dlt A_t=\veps \chi,\,\dlt A_3=-\frac{\omg}{k} \veps \chi,\, \dlt A_i=0,\\
\label{eq:ansatz2}
&\chi=\dlt A_t=\dlt A_3=0,\,\dlt A_i=\Phi^{\pm}(\rho)e^{i\omg t-ikx_3}\Y^{m,\pm}, \\
\label{eq:ansatz3}
&\chi=\dlt A_t=0,\, \dlt A_3=\Phi(\rho)e^{i\omg t-ikx_3}\Y^m,\, \dlt A_i=\frac{-ik\eta^{xx}}{m(m+2)\eta^{S3}}\Phi(\rho)e^{i\omg t-ikx_3}\nabla_i\Y^m,
\end{align}
\end{subequations}
where $\eta^{S3}$ is defined by $\eta^{ij}=\eta^{S3}\tilde{g}^{ij} = \frac{r^2}{\rho^2} \tilde{g}^{ij}$. The ansatz in eq.~\eqref{eq:ansatz1} involves only a single function $\Phi(\rho)$. The ansatz in eq.~\eqref{eq:ansatz2} involves only fluctuations with purely vector harmonics on the $S^3$, which decouple from fluctuations involving scalar harmonics since these are in different representations of $SO(4)$. In contrast, the ansatz in eq.~\eqref{eq:ansatz3} involves a coupling between $\delta A_3$, a scalar harmonic on the $S^3$, and the vector harmonics on the $S^3$ built from $\nabla_i \Y^m$. The ans\"atze for $\varphi$, $\delta A_1$, and $\delta A_2$ are the same in form to that of $\chi$ in eq.~\eqref{eq:ansatz1}, that is, $\varphi = \Phi(\rho)e^{i\omg t-ikx_3}\Y^m$, and similarly for $\delta A_1$ and $\delta A_2$.

Inserting these ans\"atze into the equations of motion, eqs.~\eqref{EOM_ex} to \eqref{eome4}, and using $A_t(\rho) = \frac{1}{(2\pi \alpha')}\frac{1}{\varepsilon} y(\rho)$ from eq.~\eqref{KO}, we find
\begin{subequations}
\begin{align}
\label{reduced_eom1}
&\del_a(\sqrt{-\eta}\eta^{ab}_S\frac{1}{\rho^2+y^2}\del_b\chi) =0,\\
\label{reduced_eom2}
&\del_k(\sqrt{-\eta}\eta^{kl}\eta^{ij}\del_j\dlt A_l)-\del_a(\sqrt{-\eta}\eta^{ij}\eta^{ac}_S \partial_c\dlt A_j)+4(\rho^2+y^2)(\rho+yy')\eps^{ijk}\del_j\dlt A_k =0,\\
\label{reduced_eom3}
&\del_a(\sqrt{-\eta}\eta^{ab}_S\frac{1}{\rho^2+y^2}\del_b\dlt A_3) =0.
\end{align}
\end{subequations}
Eqs.~\eqref{reduced_eom1}, \eqref{reduced_eom2}, and \eqref{reduced_eom3} are the non-trivial equations resulting from the ans\"atze in eqs.~\eqref{eq:ansatz1}, \eqref{eq:ansatz2}, and \eqref{eq:ansatz3}, respectively. Eqs.\eqref{reduced_eom1} and \eqref{reduced_eom3} are identical in form, and indeed are identical in form to the equations of motion of $\varphi$, $\delta A_1$, and $\delta A_2$. We have thus found that both of the scalar fluctuations and the fluctuations of the gauge field in Minkowski directions all obey the same equation of motion. Notice that these fluctuations are built from scalar harmonics on the $S^3$, \textit{i.e.} built from $\Y^m$ and/or $\nabla_i \Y^m$. The fluctuations built from genuine vector spherical harmonics $\Y^{m,\pm}$ obey a different equation of motion, eq.~\eqref{reduced_eom2}, which will actually produce two equations, for $\Phi^+(\rho)$ and $\Phi^-(\rho)$.

We thus have three ``master equations'' for the fluctuations. Let $\Phi(\rho)$ collectively denote the $\rho$-dependent part of the two scalar and the Minkowski gauge field fluctuations, and $\Phi^{\pm}(\rho)$ denote the $\rho$-dependent part of the vector harmonic fluctuations. Using the relations for the $S^3$ spherical harmonics, we can write the master equations as
\begin{subequations}
\label{master}
\begin{align}
&\del_\rho(\sqrt{-\eta}\eta^{\rho\rho}\eta^{xx}\del_\rho\Phi)-\sqrt{-\eta}\left(\omg^2\eta^{tt}\eta^{xx}+k^2(\eta^{xx})^2+m(m+2)\eta^{S3}\eta^{xx}\right)\Phi=0, \\
&\del_\rho(\sqrt{-\eta}\eta^{S3}\eta^{\rho\rho}\del_\rho\Phi^\pm)-\left(\sqrt{-\eta}\(\eta^{S3}(\omg^2\eta^{tt}+k^2\eta^{xx})+(\eta^{S3})^2(m+1)^2\)\right. \nonumber\\
&\left.\pm4(\rho^2+y^2)(\rho+yy')(m+1)\sqrt{\tilde g}\right)\Phi^\pm=0,
\end{align}
\end{subequations}
where the first equation comes from eq.~\eqref{reduced_eom1} and the second comes from eq.~\eqref{reduced_eom2}.

Recalling that the $AdS_5$ radial coordinate corresponds to the field theory energy scale, we may interpret the background solutions $y(\rho)$ and $A_t(\rho)$ as an RG flow. The fact that some of our fluctuations couple corresponds to the fact that the dual operators mix under that RG flow. For example, in eq.~\eqref{eq:ansatz1} we see that $\chi$, $\delta A_t$ and $\delta A_3$ couple to each other, but decouple from all other fluctuations, hence in the field theory the dual operators $\Om$, $J^t$, and $J^{x_3}$ mix with each other, but do not mix with any other meson operators.

To show that we have included all fluctuations, let us compare to the zero-density accounting of fluctuations of ref.~\cite{Kruczenski:2003be}. At zero density, both of the scalars decouple from all other fluctuations. As is clear from eq.~\eqref{EOM_ex}, our $\varphi$ decouples. For $\chi$ things are more subtle. Suppose we set the mass $M=0$. In that case $\veps = 0$, and from eq.~\eqref{eq:ansatz1} we see that indeed $\chi$ decouples from all other fluctuations. The reason is that when $M=0$ the chiral $U(1)_R$ symmetry is restored, and $\varphi$ and $\chi$ are the only worldvolume fields charged under that symmetry \cite{Kruczenski:2003be}. In other words, they are in a different representation of $U(1)_R$ from all the other fluctuations, and hence must decouple. When $M$ is nonzero, the $U(1)_R$ is broken. Nothing then forbids $\chi$ from coupling to other fluctuations, so we must allow for fluctuations of the form in eq.~\eqref{eq:ansatz1}, although in the end $\chi$ does decouple, as we see in eq.~\eqref{reduced_eom1}. The upshot is that eq.~\eqref{eq:ansatz1} is the generalization to finite density of the $\chi$ fluctuation in ref.~\cite{Kruczenski:2003be}.

In ref.~\cite{Kruczenski:2003be}, three types of gauge field fluctuations were identified, which were called type I, II, and III. Type I were identical to our eq.~\eqref{eq:ansatz2}, and our eq.~\eqref{eq:ansatz3} is essentially the generalization of the type III fluctuations to finite density and to our choice of gauge. The type II fluctuations involved $S^3$ scalar harmonics of the Minkowski components of the gauge field, which decoupled from all other fluctuations. In our case, where the charge density breaks Lorentz symmetry to spatial rotations and translations, the Minkowski components of the gauge field split: $\delta A_1$ and $\delta A_2$ still decouple, but as we discussed above, when $M$ is nonzero $\delta A_t$ and $\delta A_3$ mix with $\chi$.

Notice that $m=0$ is a special case. When $m=0$, the ansatz in eq.~\eqref{eq:ansatz3} is ill-defined. That is not unexpected, since the type III fluctuations of ref.~\cite{Kruczenski:2003be} were also singular when $m=0$. For the ansatz in eq.~\eqref{eq:ansatz1}, eqs.~\eqref{eome1} and \eqref{eome4} are constraint equations, coming from setting $\delta A_{\rho} = 0$ and $\delta A_i = 0$, respectively. When $m=0$, eq.~\eqref{eome4} is trivially satisfied, meaning one constraint is eliminated. We then expect to find an additional degree of freedom, which in our case is the fluctuation giving rise to the zero sound mode, as we review in the appendix.

We will make two final simplifications, to make an analysis of the singular points of the equations of motion easier, and also to make our numerical analysis easier. First, we will define dimensionless variables. Let $\bar{N} \equiv \N/c$, which has length dimension three (recall eq.~\eqref{eq:cddefs}), which we use to define the dimensionless variables $\brho$, $\by$, $\bomg$, and $\bk$:
\be
\label{eq:rescaledvariables}
&&\brho^6\equiv \rho^6 \bN^2 \left(\frac{1}{\veps^2}-1\right)^{-1}, \qquad \by^6 \equiv y^6 \bN^2 \left(\frac{1}{\veps^2}-1\right)^{-1}, \\
&&\bomg^2 \equiv \omg^2\(\bN^2/\left(\frac{1}{\veps^2}-1\right)\)^{1/3}, \qquad \bk^2 \equiv k^2\(\bN^2/\(\frac{1}{\veps^2}-1\)\)^{1/3}. \nonumber
\ee
In terms of these new variables, we have
\be
\label{eq:rescaledysol}
\by(\brho)=\frac{1}{6}\(\frac{1}{\veps^2}-1\)^{-1/2}B\left(\frac{\brho^6}{\brho^6+1};\frac{1}{6},\frac{1}{3}\right), \quad \by'(\brho)\equiv\frac{d\by}{d\brho}=\frac{1}{\sqrt{(\brho^6+1)\(\frac{1}{\veps^2}-1\)}}.
\ee
The three master equations now take the form
\begin{subequations}
\begin{align}
\label{phiscalar}
&\Phi''+p_1(\brho)\Phi'+q_1(\brho,\bomg,\bk,m)\Phi=0, \\
\label{phivector}
&\Phi^\pm{}''+p_2(\brho)\Phi^\pm{}'+q_2^{\pm}(\brho,\bomg,\bk,m)\Phi^\pm=0,
\end{align}
\end{subequations}
where
\begin{subequations}
\begin{align}
\label{eq:p1def}
&p_1(\brho)=\frac{3\brho^5}{1+\brho^6}, \\
\label{eq:q1def}
&q_1(\brho,\bomg,\bk,m)=\frac{-m(m+2)\brho^4}{1+\brho^6}+\frac{\bomg^2(1+\by'^2)}{(\brho^2+\by^2)^2}-\frac{\bk^2\brho^6}{(1+\brho^6)(\brho^2+\by^2)^2}, \\
\label{eq:p2def}
&p_2(\brho)=\frac{\brho^2(2+5\brho^6)+(-2+\brho^6)\by^2+4\brho(1+\brho^6)\by\by'}{\brho(1+\brho^6)(\brho^2+\by^2)}, \\
\label{eq:q2def}
&q_2^{\pm}(\brho,\bomg,\bk,m)=-\frac{(m+1)^2\brho^4}{1+\brho^6}\mp4(m+1)\sqrt{\frac{\brho^6}{1+\brho^6}}\frac{\brho+\by\by'}{\brho(\brho^2+\by^2)}\\
&\qquad \qquad \qquad +\frac{\bomg^2(1+\by'^2)}{(\brho^2+\by^2)^2}-\frac{\bk^2\brho^6}{(1+\brho^6)(\brho^2+\by^2)^2}. \nonumber
\end{align}
\end{subequations}
The master equations, eqs.~\eqref{phiscalar} and \eqref{phivector}, and their solutions, will be the focus of the remainder of this paper.

Finally, let us demonstrate how the fluctuations ``see'' a near-horizon $AdS_2$ region \cite{Jensen:2010ga,Jensen:2010vx, Evans:2010np,Nickel:2010pr}. In our case ``near-horizon'' means $\brho \to 0$. In that limit we have
\be
\by(\brho) = \frac{\brho}{\sqrt{\frac{1}{\veps^2} -1}} + O\left( \brho^7\right).
\ee
Consider the equation for $\Phi(\brho)$, eq.~\eqref{phiscalar}. From eq.~\eqref{eq:q1def} we see that as $\brho \to 0$, the dominant term in the coefficient $q_1(\brho,\bomg,\bk,m)$ is the one proportional to $\bomg^2$, which will scale at leading order as $1/\brho^4$, while the term proportional to $m(m+2)$ scales as $\brho^4$ and the term proportional to $\bk^2$ scales as $\brho^6$. Here we already see a hint of $AdS_2$: only the radial and time directions will appear in $AdS_2$, so we should see a suppression of terms involving spatial momenta. Here we actually see a suppression of terms involving the linear momentum in Minkowski directions as well as the angular momentum in the $S^3$ directions. From eq.~\eqref{eq:p1def} we can see that when $\brho \to 0$ the coefficient $p_1(\brho) \to 3\brho^5$, which we will treat as sub-leading, \textit{i.e.} we will drop the $\Phi'$ term from eq.~\eqref{phiscalar} in the near-horizon limit. The equation for $\Phi(\brho)$ in the $\brho \to 0$ limit thus becomes, at leading order,
\be
\label{eq:nearhorizonphiscalar}
\Phi'' + \frac{\bomg^2}{\brho^4} (1-\veps^2) \Phi =0. 
\ee
If we extract a factor of $\brho$, that is, if we define $\Phi(\brho) \equiv \brho \hat{\Phi}(\brho)$, then eq.~\eqref{eq:nearhorizonphiscalar} becomes
\be
\label{eq:nearhorizonphiscalarrescaled}
\hat{\Phi}'' + \frac{2}{\brho} \hat{\Phi}' + \frac{\bomg^2}{\brho^4} (1-\veps^2) \hat{\Phi}=0,
\ee
which is precisely the equation of motion for a massless scalar in $AdS_2$, with metric $\frac{d\brho^2}{\brho^2} - \brho^2 dt^2$, and with frequency $\bomg \sqrt{1 - \veps^2}$.

The $\brho \to 0$ limit for the $\Phi^{\pm}(\brho)$ equation, eq.~\eqref{phivector}, is even simpler. From eq.~\eqref{eq:q2def} we see that as $\brho \to 0$ the leading term in $q_2(\brho,\bomg,\bk,m)$ is identical to the leading term in $q_1(\brho,\bomg,\bk,m)$, namely $\bomg^2 (1-\veps^2)/\brho^4$. From eq.~\eqref{eq:p2def} we find that $p_2(\brho)$ has an especially simple behavior as $\brho \to 0$, namely $p_2(\brho) \to 2/\brho$. The equation for the $\Phi^{\pm}(\brho)$ in the $\brho \to 0$ limit thus becomes, to leading order,
\be
\label{eq:nearhorizonphivector}
\Phi^\pm{}''+\frac{2}{\brho}\Phi^\pm{}'+ \frac{\bomg^2}{\brho^4} (1-\veps^2)\Phi^\pm=0,
\ee
which again is precisely the equation of motion for a massless scalar in $AdS_2$ with frequency $\bomg \sqrt{1 - \veps^2}$. Here we did not even need to rescale $\Phi^{\pm}(\brho)$ by a factor of $\brho$, in contrast to $\Phi(\brho)$.

We emphasize that the appearance of a near-horizon $AdS_2$ region is due to the presence of a nontrivial $A_t(\brho)$, or in field theory language a finite density. At zero density, Lorentz invariance demands that the equations of motion depend only on the Lorentz-invariant quantity $\omega^2 - k^2$, hence dependence on $k$ could not be suppressed relative to $\omega$ in the near-horizon limit.

Given that eqs.~\eqref{eq:nearhorizonphiscalarrescaled} and \eqref{eq:nearhorizonphivector} describe a \textit{massless} scalar field in $AdS_2$, for all values of $\bk$, $\veps$, and $m$, we will not see an instability involving a violation of the $AdS_2$ BF bound, that is, an instability arising from the mass-squared of a field in $AdS_2$ becoming too negative. Of course, many other kinds of instabilities are possible. In fact, for our system two deformations are known to produce instabilities without any violation of the $AdS_2$ BF bound. The first is a sufficiently large $U(1)_B$ magnetic field, which for massless flavors causes spontaneous breaking of the chiral $U(1)_R$ symmetry \cite{Filev:2007gb,Evans:2010iy,Jensen:2010vd}. The second is an isospin chemical potential, meaning we introduce a second D7-brane coincident with the first, so that the field theory flavor symmetry is enhanced to $U(1)_B \times SU(2)$, and we then introduce a chemical potential associated with the Cartan generator of $SU(2)$. That causes vector meson condensation, leading to a p-wave superfluid state \cite{Ammon:2008fc,Basu:2008bh,Ammon:2009fe,Erdmenger:2011hp}. In each case a near-horizon $AdS_2$ is present, but no violation of the $AdS_2$ BF bound occurs.

\section{The Spectrum of Quasi-Normal Modes}\label{ss:spectrum}

Every fluctuation of a D7-brane worldvolume field is dual to a meson operator. The AdS/CFT dictionary equates the on-shell bulk action with the field theory generating functional. The solutions of the bulk linearized equations of motion, eqs.~\eqref{phiscalar} and \eqref{phivector} thus determine the dual two-point functions. In Lorentzian signature we have a choice of two-point functions: retarded, advanced, etc. In the bulk these are determined by the boundary condition we impose in the near-horizon region. We want the retarded two-point functions, so we will impose in-going boundary conditions in the near-horizon region, \textit{i.e.} the solutions will look like waves traveling towards the bottom of $AdS_5$ \cite{Son:2002sd}.

We will not compute the full two-point functions. To study stability, simply computing the eigenfrequencies of the linear operators in eqs.~\eqref{phiscalar} and \eqref{phivector}, that is, the QNM's of the system, will be sufficient. These QNM's are dual to poles in the retarded Green's functions. An instability would appear as a pole in the lower half of the complex frequency plane\footnote{Recall again our sign convention, such that stable modes appear in the upper-half of the complex frequency plane.}, since such a mode would have uncontrolled exponential growth.

The main technical difficulty in computing the QNM's is the presence of irregular singular points in eqs.~\eqref{phiscalar} and \eqref{phivector}. To deal with these, we will employ a combination of analytical and numerical techniques. In section~\ref{lowfreq} we consider the low-frequency limit, where we can match asymptotic expansions to determine analytically the low-frequency form of the two-point function, following ref.~\cite{Faulkner:2009wj}. In section~\ref{ss:zigzag} we discuss the singular points of eqs.~\eqref{phiscalar} and \eqref{phivector}, review Leaver's method for dealing with these, explain why Leaver's method is difficult to use in our case, and then present our own numerical method, which we use to compute the QNM spectra presented in section~\ref{numerical}.

\subsection{Low-Frequency Expansion} \label{lowfreq}

We expect an instability to appear as a QNM crossing into the lower-half of the complex frequency plane at small frequency. Luckily, we can obtain the exact form of the retarded Green's functions in the low-frequency regime, following ref.~\cite{Faulkner:2009wj}. We divide the radial direction into two regions, the ``inner'' and ``outer'' regions, defined by $\brho \ll 1$ and $\Omega/\brho \ll 1$, respectively. These regions overlap at small frequency $\Omega \ll 1$. The inner region is simply the near-horizon $AdS_2$ region.

The presence of the near-horizon $AdS_2$ suggests that in the field theory in these finite-density states some $(0+1)$-dimensional CFT emerges at low energies. We may invoke an AdS/CFT correspondence for the $AdS_2$ region itself: every field in $AdS_2$ will be dual to some operator in the dual $(0+1)$-dimensional CFT, living at the boundary of $AdS_2$, where we match to the outer region. For example, if a bulk field in the $AdS_2$ region obeys the equation of a massless scalar, then that field is dual to some marginal (dimension one) scalar operator in the $(0+1)$-dimensional CFT. We expect the Green's functions of the $(0+1)$-dimensional CFT to determine the low-frequency behavior of the full Green's functions \cite{Faulkner:2009wj},  which indeed occurs, as we will show.

In the last section we saw that \textit{all bosonic} fluctuations of D7-brane fields ultimately obey the same equation of motion in the $AdS_2$ region, that for a massless scalar, regardless of their spin (scalar or vector), their linear momentum in Minkowski directions, or their angular momentum in $S^3$ directions. We thus conclude that in the field theory, every bosonic meson operator is dual at low energies to some marginal scalar operator in the $(0+1)$-dimensional CFT, regardless of its spin, momentum, or its $SO(4)$ charges.

We begin with $\Phi(\brho)$. The equation of motion in the inner region is eq.~\eqref{eq:nearhorizonphiscalar}. Extracting a factor of $\brho$ via $\Phi(\brho) = \brho \hat{\Phi}(\brho)$, we obtain eq.~\eqref{eq:nearhorizonphiscalarrescaled} for a massless scalar in $AdS_2$. The in-going solution to eq.~\eqref{eq:nearhorizonphiscalarrescaled} is
\be
\label{eq:innerregionsol}
\hat{\Phi}(\brho) = e^{-\frac{i\Omg}{\brho}} = 1 - \frac{i \Omg}{\brho} + {\cal O}\left( \frac{\Omg^2}{\brho^2}\right), \qquad \Omega \equiv \bomg \sqrt{1-\veps^2},
\ee
where in the second equality we have expanded in $\Omega/\brho \ll 1$. The leading terms are a constant and $\brho^{-1}$, precisely as we expect for a massless scalar in $AdS_2$, dual to a dimension-one operator. Following the standard AdS/CFT recipe, we identify the coefficient of the leading, non-normalizable term, the constant, as the source for the dual operator, and the coefficient of the sub-leading normalizable term, $\brho^{-1}$, as the one-point function. The dual two-point function is thus the ratio of these coefficients, $-i \Omg$, which is the expected result \cite{Faulkner:2009wj}. The full inner-region solution for $\Phi(\brho)$, including corrections, is
\be
\Phi(\brho)=\brho \, e^{-\frac{i\Omg}{\brho}}(1+{\cal O}(\brho^5)).
\ee
At low frequency we can approximate the solution as
\be
\label{eq:innerregionsolapprox}
\Phi(\brho) \approx \brho - i \Omg.
\ee

In the outer region we obtain the leading-order small-frequency solution from eq.~\eqref{phiscalar} by setting $\bomg=0$. The resulting equation will have two linearly independent solutions which we will denote $\Phi_0(\brho)$ and $\Phi_1(\brho)$. These will depend on the parameters $(\bk,\veps,m)$ but obviously not on $\Omg$. We then construct the general solution by taking a linear combination of $\Phi_0(\brho)$ and $\Phi_1(\brho)$, with coefficients that depend not only on $(\bk,\veps,m)$, but also on $\Omg$. To match to the inner region, we demand that $\Phi_0(\brho)$ approach a constant, say one, as $\brho \to 0$, and that $\Phi_1(\brho)$ approach $\brho$ as $\brho \to 0$. That fixes the $\Omg$-dependence of the ratio of the coefficients. Explicitly, we find at low frequency
\be
\Phi(\brho) = \Phi_1(\brho) - i \Omg \Phi_0(\brho).
\ee
Near the $AdS_5$ boundary we will find two solutions, one normalizable and one not. We denote these $\Phi_V(\brho)$ and $\Phi_S(\brho)$, respectively, where ``V'' and ``S'' stand for ``vacuum expectation value'' and ``source.'' These will be linearly related to $\Phi_1(\brho)$ and $\Phi_0(\brho)$:
\begin{subequations}
\be
&&\Phi_1(\brho)=a_1\Phi_S(\brho)+b_1\Phi_V(\brho), \\
&&\Phi_0(\brho)=a_0\Phi_S(\brho)+b_0\Phi_V(\brho),
\ee
\end{subequations}
where we have suppressed the $(\bk,\veps,m)$-dependence of the coefficients $a_1$, $b_1$, $a_0$, $b_0$. Near the $AdS_5$ boundary we thus have
\be
\Phi(\brho) = (a_1 - i\Omg a_0) \Phi_S(\brho) + (b_1 - i\Omg b_0) \Phi_V(\brho).
\ee
The coefficient of the second term is the one-point function and the coefficient of the first term is the source, hence the retarded Green's function, $G(\Omg)$, is simply the ratio 
\be
\label{somgG}
G(\Omg)=\frac{b_1-i\Omg b_0}{a_1-i\Omg a_0}.
\ee
Notice that we have suppressed $G(\Omg)$'s dependence on $(\bk,\veps,m)$. Eq.~\eqref{somgG} is the leading low-frequency form of the retarded Green's function for all operators dual to fluctuations obeying eq.~\eqref{phiscalar}. Clearly the Green's function of the (0+1)-dimensional CFT, $-i \Omg$, determines the low-frequency behavior of the full Green's function.

For $\Phi^{\pm}(\brho)$, the analysis is nearly identical. The main difference is that in the inner region we do not extract a factor of $\brho$ from $\Phi^{\pm}(\brho)$, so eq.~\eqref{eq:innerregionsol} is the inner region solution rather than eq.~\eqref{eq:innerregionsolapprox}, or in other words in the inner region, $\brho$ and a constant are replaced by a constant and $\brho^{-1}$, respectively. We denote the outer region solutions that match onto these as $\Phi^{\pm}_0(\brho)$ and $\Phi^{\pm}_{-1}(\brho)$. The rest of the procedure is unchanged. In the outer region $\Phi^{\pm}_0(\brho)$ and $\Phi^{\pm}_{-1}(\brho)$ are linearly related to the normalizable and non-normalizable solutions as (in hopefully obvious notation)
\begin{subequations}
\be
&&\Phi_0^\pm(\brho)=a_0^\pm\Phi_S^\pm(\brho)+b_0^\pm\Phi_V^\pm(\brho), \\
&&\Phi_{-1}^\pm(\brho)=a_{-1}^\pm\Phi_S^\pm(\brho)+b_{-1}^\pm\Phi_V^\pm(\brho).
\ee
\end{subequations}
and the retarded Green's function at low frequencies is
\be\label{somgGpm}
G^\pm(\Omg)=\frac{b_0^\pm-i\Omg b_{-1}^\pm}{a_0^\pm-i\Omg a_{-1}^\pm}.
\ee

The coefficients $a_1,\,a_0,\,b_1,\,b_0$ and $a^{\pm}_0,\,a^{\pm}_{-1},\,b^{\pm}_0,\,b^{\pm}_{-1}$ are necessarily real, as is clear from the zero-frequency form of eqs.~\eqref{phiscalar} and \eqref{phivector}. To obtain them requires solving for $\Phi(\brho)$ and $\Phi(\brho)^{\pm}$ for all $\brho$, which we have only been able to do numerically. Crucially, however, to solve for the coefficients we can set $\Omg =0$, in which case $\brho=0$ is a \textit{regular} singular point of eqs.~\eqref{phiscalar} and \eqref{phivector}, for which we need no special methods to solve the equations. We may thus straightforwardly compute these coefficients for various values of $(\bk,\veps,m)$ and, using eqs.~\eqref{somgG} and \eqref{somgGpm}, we may look for instabilities. A QNM with negative imaginary part would appear when $\frac{a_1}{a_0}\to 0^-$ or $\frac{a_0^{\pm}}{a_{-1}^{\pm}}\to 0^-$.

We have studied numerically the $m=0,1$ modes of $\Phi(\brho)$, the $m=1$ mode of $\Phi^+(\brho)$, and the $m=1,2,3$ modes of $\Phi^-(\brho)$ for values of $\veps \in (0,0.9)$ and $\bk \in (0,50)$. For $\Phi(\brho)$ and $\Phi^+(\brho)$, we detected no instabilities. For $\Phi^-(\brho)$ we found that $\frac{a_0^-}{a_{-1}^-}\to D(\veps) \bk^2>0$ as $\bk\to 0$, for some function $D(\veps)$ that we can compute numerically (see figure~\ref{fig:diffusion}). Such behavior does not indicate an unstable mode, but rather a mode with a dispersion relation $\Omg=i D(\veps)\bk^2$. This is our R-spin diffusion mode. The operator dual to $\Phi^-(\brho)$ with $m=1$  is a dimension-two Lorentz scalar in the vector representation of $SU(2)_R$ \cite{Kruczenski:2003be}. To be explicit, let $Q^{\alpha} = (q,\tilde{q}^{\dagger})^T$ denote the $SU(2)_R$ doublet of squarks. $\Phi^-(\brho)$ with $m=1$ is then dual to
\be
\label{eq:oidef}
{\cal O}^I = Q^{\alpha}{}^{\dagger}\sigma^I_{\alpha \beta}Q^{\beta},
\ee
where $\sigma^I_{\alpha \beta}$ are the $SU(2)_R$ Pauli matrices.

We mostly consider modes with $m \leq 2$. We can make a heuristic argument for why modes with higher $m$ are unlikely to exhibit instabilities. Recall from section~\ref{ss:theory} that the value of $m$ determines the dimension of the dual operator. Larger $m$ means larger operator dimension. From a field theory point of view, we expect instabilities to appear first in the most relevant operators, \textit{i.e.} the lowest-dimension operators, as we cool the system through a phase transition, for example. If we see no instabilities for small $m$, then we have good reason to expect that no instabilities will appear for larger $m$.

\subsection{The Zig-Zag Method}\label{ss:zigzag}

We now come to the problem of computing the QNM's of our system numerically. The main obstacle is the presence of irregular singular points in eqs.~\eqref{phiscalar} and \eqref{phivector}. In particular, $\brho=0$ is a simple pole of the coefficients $p_1(\brho)$ and $p_2(\brho)$ but a fourth-order pole of $q_1(\brho,\bomg,\bk,m)$ and $q_2(\brho,\bomg,\bk,m)$

Let us illustrate the problem using the near-horizon equation for $\Phi(\brho)$, eq.~\eqref{eq:nearhorizonphiscalar}. That equation has two solutions, $\brho e^{\pm i \Omg /\brho}$, where the minus sign gives the in-going solution. If we introduce a cutoff at some small but finite $\brho$, impose the in-going wave boundary condition, and numerically integrate the equation of motion, the results are extremely sensitive to changes in the cutoff, and are thus unphysical. Moreover, since the QNM usually contains a non-vanishing imaginary part, the in-going solution is exponentially growing as $\brho \to 0$ while the out-going solution is exponentially suppressed. Since our series solution $\brho e^{-\frac{i\Omg}{\brho}}(1+{\cal O}(\brho^5))$ is only accurate sufficiently close to the horizon, numerical error will always effectively ``include'' the exponentially suppressed solution, leading to an inaccurate result.

A few techniques have been developed to deal with such problems \cite{Leaver:1990zz,Nunez:2003eq,Faulkner:2009wj,Denef:2009yy,Kaminski:2009ce}. One is to use matched asymptotic expansions at low frequency, which we employed above. Another technique, valid for any frequency, is the matrix method, or Leaver's method \cite{Leaver:1990zz,Denef:2009yy}. Consider $\Phi(\brho)$ as an example. With Leaver's method we begin by extracting from $\Phi(\brho)$ the leading behavior near each singular point, so we set $\Phi(\brho) = \brho e^{-i \Omg/\brho} \tilde{\Phi}(\brho)$. Inserting that into eq.~\eqref{phiscalar}, we obtain an equation for $\tilde{\Phi}(\brho)$. To solve that equation, we choose a non-singular point of the equation, $\tilde{\rho}$, such that when we expand $\tilde{\Phi}(\brho)$ about that point the radius of convergence includes the two singular points at the horizon and at the boundary,
\be
\label{eq:leaverexpansion}
\tilde{\Phi}(\brho) = \sum_{p=0}^n c_p (\brho - \tilde{\rho})^p,
\ee
with some coefficients $c_p$ that depend on $\bomg$, $\bk$, $\veps$, and $m$. We truncate the series at some order $n$ and then insert eq.~\eqref{eq:leaverexpansion} into the equation of motion. We obtain $n+1$ equations for the $n+1$ unknown coefficients, which we can assemble into a matrix equation,
\be
\sum_{p=0}^n {\cal M}_{qp} c_p = 0,
\ee
where ${\cal M}_{qp}$ also depends on $\bomg$, $\bk$, $\veps$, and $m$. We then find the QNM's by finding where $\det {\cal M} = 0$. Upon increasing $n$, the results should converge to the physical values.

Leaver's method works well for fields in extremal AdS-RN \cite{Denef:2009yy,Edalati:2010hk,Edalati:2010pn}. For example, for a charged scalar field \cite{Denef:2009yy}, the coefficients $c_p$ obey a nine-term recurrence relation, so that ${\cal M}_{qp}$ only has nonzero entries along a diagonal band of width nine. A numerical calculation of $\det {\cal M}$ is then reasonably efficient, even for large values of $n$.

Leaver's method is more difficult to use in our case. In general, we found that ${\cal M}$ is upper triangular, indicating that the $c_p$ obey a more complicated recurrence relation compared to an AdS-RN charged scalar. The calculation of ${\cal M}$ becomes more and more difficult as we increase $n$, since we have to compute the determinant of a large upper triangular matrix rather than of a reasonably sparse matrix with entries only along a diagonal band.

To make progress, we developed our own method, which works as follows. First we pick some value $\bomg$. We want to test whether that value is a QNM, so we denote it $\bomg^{\text{test}}_{\text{QNM}}$. Next we analytically continue $\brho$ to complex values, such that $\arg\brho=\arg\bomg^{\text{test}}_{\text{QNM}}$. The benefit of doing so is that in the exponent of $\brho e^{\pm i\Omg/\brho}$, the ratio $\Omg/\brho$ is now purely real, so the in-going and out-going solutions are purely oscillatory and thus easy to disentangle. Now we just need to specify a contour for the $\brho$ integration that takes us from $\brho=0$ to the point at infinity on the real $\brho$ axis, starting along the line $\arg\bomg^{\text{test}}_{\text{QNM}}$ for small values of $|\brho|$.

Complexifying $\brho$ comes at a cost, however, since branch cuts appear in the complex $\brho$ plane. These arise from the various factors of $\by(\brho)$ in the coefficients $p_2(\brho)$, $q_1(\brho,\bomg,\bk,m)$, and $q_2(\brho,\bomg,\bk,m)$. To see these, recall from eq.~\eqref{eq:rescaledysol} that $\by(\brho)$ is an incomplete Beta function, which can be expressed in terms of a hypergeometric function. In our case, for real $\brho$ we have
\be
B\left(\frac{\brho^6}{\brho^6+1};\frac{1}{6},\frac{1}{3}\right) = \frac{6 \brho}{(1+\brho^6)^{1/6}} \, _2F_1\left(\frac{1}{6},\frac{2}{3};\frac{7}{6};\frac{\brho^6}{1+\brho^6}\right).
\ee
The hypergeometric function has a branch cut for values of its argument from one to infinity. Upon complexifying $\brho$, in our case these points correspond to $\brho =\infty$ and the points $\brho^6 + 1=0$, respectively. We thus find six branch cuts in the complex $\brho$ plane, as illustrated in figure~\ref{zigzag}.

\FIGURE[t]{\includegraphics[width=9cm]{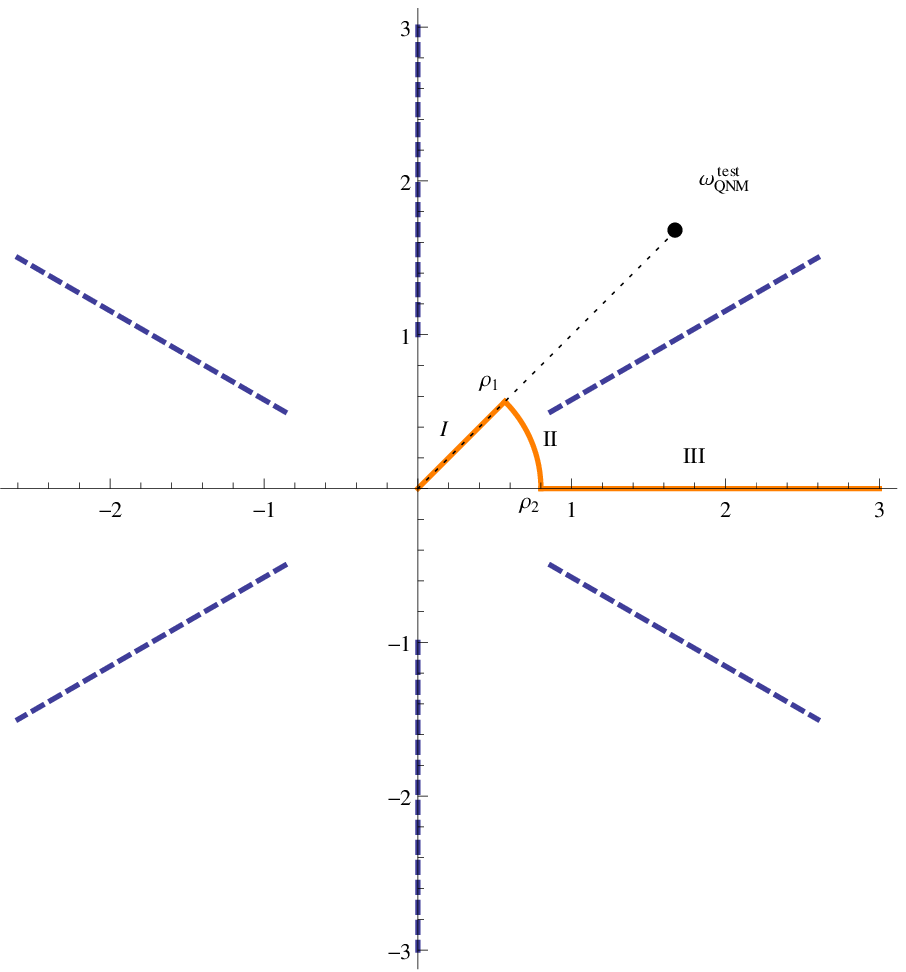}
\caption{(Color online) The complex $\brho$ plane, with the branch cuts in the coefficients of eqs.~\eqref{phiscalar} and \eqref{phivector} indicated by dashed blue lines and our contour of integration indicated by the solid orange curve, with segments I, II and III labeled, as well as the turning points $\brho_1$ and $\brho_2$. The dotted black line ending in the dot labeled $\bomg^{\text{test}}_{\text{QNM}}$ indicates $\arg\bomg^{\text{test}}_{\text{QNM}}$.}
\label{zigzag}}

The integration of the equations of motion in the complex $\brho$ plane will obviously encounter problems if $\brho$ crosses these branch cuts. We thus employ the ``zig-zag'' integration contour illustrated in figure~\ref{zigzag}. For small $|\brho|$ we move along the line $\arg\bomg^{\text{test}}_{\text{QNM}}$. For some $|\brho|$ large enough that the contamination from the out-going solution is negligible, but smaller than the point where the branch cut begins, $|\brho|=1$, we turn toward the real axis, moving along an arc at the fixed radius $|\brho|$, until we reach $\arg \brho =0$. We then continue integrating along the real axis to $\brho = \infty$, or actually to some large cutoff value.

We must specify matching conditions at each of the turning points of the contour. Let us denote the two turning points as $\brho_1$ and $\brho_2$, and the value of $\Phi(\brho)$ on the three pieces of the contour as $\Phi_I(\brho)$, $\Phi_{II}(\brho)$, and $\Phi_{III}(\brho)$ (see figure~\ref{zigzag}). To guarantee that solutions will be analytic in the complex $\brho$ plane (away from the branch cuts), we require that the function and its first derivative agree at the turning points:
\begin{subequations}
\be
&&\Phi_I(\brho_1)=\Phi_{II}(\brho_1),\qquad\Phi_I'(\brho_1)=\Phi_{II}'(\brho_1), \\
&&\Phi_{II}(\brho_2)=\Phi_{III}(\brho_2),\qquad\Phi_{II}'(\brho_2)=\Phi_{III}'(\brho_2),
\ee
\end{subequations}
where the derivatives are taken along the corresponding piece of the contour.

To illustrate our procedure from start to finish, consider for example $\Phi(\brho)$. We first expand $\Phi(\brho)$ near the horizon,
\be
\Phi(\brho)=e^{-i\Omg/\brho}\left(\brho+\sum_{p=6}^\infty \alpha_p \, \brho^p\right),\nonumber
\ee
where we calculate the constant coefficients recursively from $\alpha_6$ up to $\alpha_{p_{\textrm{max}}}$, with $p_{\textrm{max}}$ typically of order twelve. For the numerical integration, we typically imposed a small-$\brho$ cutoff of $\brho\leq0.1$, although the results of the zig-zag method were extremely stable against changes in the cutoff. We then integrate the equation numerically along the zig-zag contour up to some large-$\brho$ cutoff, near the $AdS_5$ boundary. We then fit the solution to the leading normalizable and non-normalizable solutions in the large-$\brho$ region (explicitly restoring $\Phi(\brho)$'s dependence on $\bomg$ and $\bk$ for the moment),
\be
\Phi(\brho,\bomg,\bk) \approx A(\bomg,\bk) \Phi_S(\brho) + B(\bomg,\bk) \Phi_V(\brho).
\ee
For example, when $m=0$,\footnote{For general $m$, at leading order $\Phi_S(\brho) \approx \brho^m$ and $\Phi_V(\brho) \approx \brho^{-m-2}$. The leading behavior of $\Phi_V(\brho)$ agrees with that obtained for the scalar and type II and III gauge fluctuations in ref.~\cite{Kruczenski:2003be}, as expected. For $\Phi^{\pm}(\brho)$, at leading order $\Phi_S^+(\brho)=\brho^{m+1}$ and $\Phi_V^+(\brho)=\brho^{-m-5}$, while $\Phi_S^-(\brho)=\brho^{m-3}$ and $\Phi_V^-(\brho)=\brho^{-m-1}$. The leading behaviors of $\Phi^{\pm}_V(\brho)$ agree with those of the type I fluctuations in ref.~\cite{Kruczenski:2003be}.}
\be
\Phi_S(\brho) = 1+\frac{(\bomg^2-\bk^2)\log{\brho}}{2\brho^2}+{\cal O}(\brho^{-3}), \quad \Phi_V(\brho)=\brho^{-2}+{\cal O}(\brho^{-3}).
\ee
The coefficient $A(\bomg,\bk)$ acts as the source for the dual operator, while $B(\bomg,\bk)$ is the one-point function. The retarded Green's function, in the regime of linear response, is thus (proportional to) the ratio $B(\bomg,\bk)/A(\bomg,\bk)$. To locate the QNM's we looked for zeroes of $|A(\bomg,\bk)/B(\bomg,\bk)|$. We performed the search for the zeroes numerically using our own two-dimensional (over the complex $\bomg$ plane) minimization algorithm.\footnote{A Mathematica notebook implementing the minimization algorithm is available from J.~Shock upon request.} Actually, in a numerical procedure $|A(\bomg,\bk)/B(\bomg,\bk)|$ will never be exactly zero, so we searched for values of $|A(\bomg,\bk)/B(\bomg,\bk)|$ below some cutoff typically between $10^{-4}$ and $10^{-2}$. Our procedure for $\Phi^{\pm}(\brho)$ was very similar to that of $\Phi(\brho)$.

As mentioned above, in all cases the results of zig-zag method were extremely stable against changes in the small-$\brho$ cutoff. As a more rigorous test, we compared the results of the zig-zag method to the low-frequency forms of the two-point functions derived in section~\ref{lowfreq}. Since we only computed QNM's, and not the full two-point functions, we compared only to $G^-(\Omg)$ in eq.~\eqref{somgGpm} because that has a QNM at low-frequency, the R-spin diffusion mode with dispersion relation $\Omg = i D(\veps) k^2$. The results of the zig-zag method, in the low-frequency limit, produced precisely this dispersion relation for the R-spin diffusion mode.

\subsection{Numerical Results}\label{numerical}

For $\Phi(\brho)$ and $m=0$, the first three QNM's appear in figure~\ref{fig:Phi0QNMs}, for $\veps$ up to order one and $\bk$ up to order ten. For larger $\bk$ we were unable to identify zeroes of $|A(\bomg,\bk)/B(\bomg,\bk)|$ numerically because the value of $|A(\bomg,\bk)/B(\bomg,\bk)|$ in the immediate neighborhood of the zero is larger by orders of magnitude, \textit{i.e.} numerically we find a large ``background'' surrounding the zero, so that identifying the exact location of the zero is difficult. Notice that the QNM's are mirror-symmetric about the imaginary $\bomg$ axis.

\FIGURE[t]{\includegraphics[width=10cm]{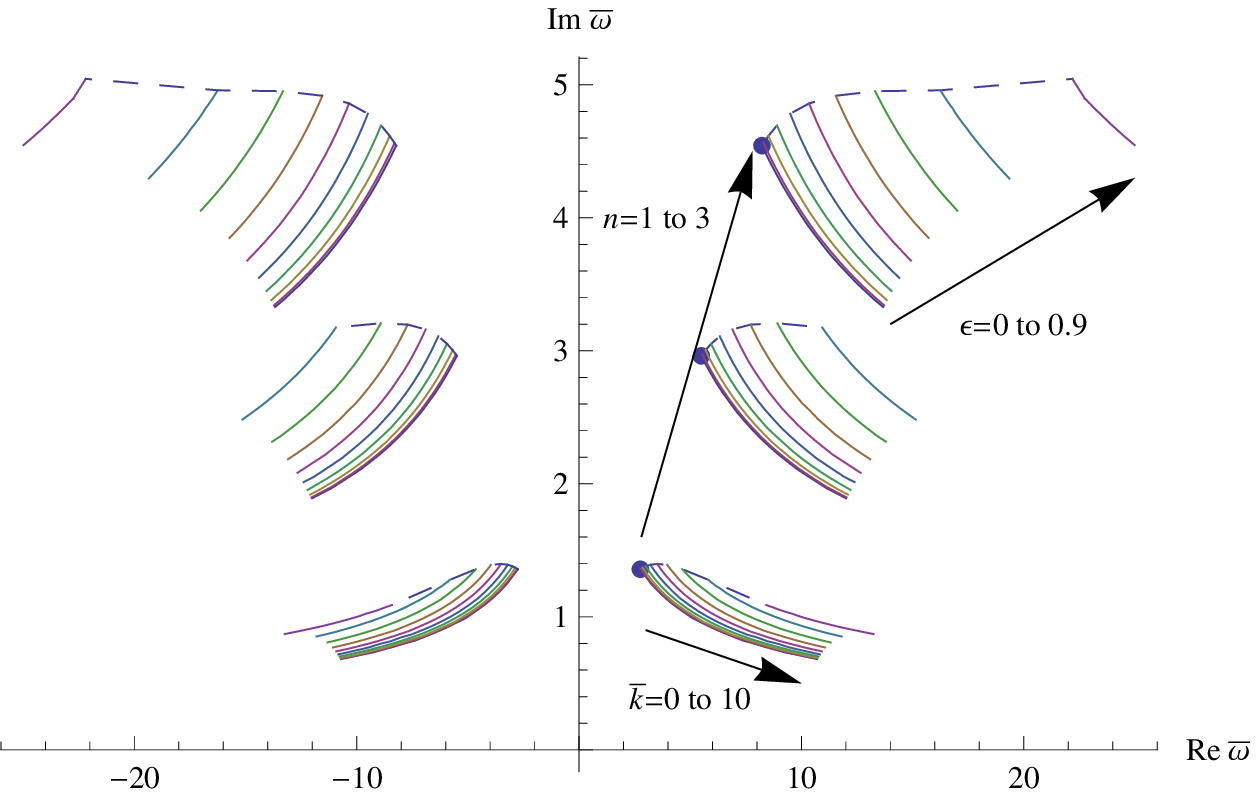}
\caption{(Color online) The complex $\bomg$ plane, where the curves show the motion of the first three QNM's of $\Phi(\brho)$ for $m=0$, labeled by $n=1,2,3$. The solid lines describe the motion as a function of momentum $\bk \in (0,10)$ while the ten lines for each $n$ (nine for $n=2$) are for $\veps= M/\mu = 0$ to $0.9$ (0.8 for $n=2$) in steps of 0.1. The dashed lines indicate the behavior for $\bk=0$ as we increase $\veps$. None of the modes cross into the lower half-plane for our ranges of $\veps$ and $\bk$, hence none become unstable.}
\label{fig:Phi0QNMs}}

None of the modes that we see ever cross into the lower half of the complex $\bomg$ plane, that is, none become unstable. One unusual feature we do see from figure~\ref{fig:Phi0QNMs} is that the imaginary part of the QNM decreases as the momentum $\bk$ increases. (Contrast that with the zero sound dispersion relation in eq.~\eqref{eq:zsdispersion}.) Figure~\ref{fig:asymvel} shows the real part of the first QNM as function of $\bk$, for $\veps \in (0,0.9)$. At large $\bk$ we see that the modes are approaching the lightcone. Numerically we find that by $\bk \approx 10$ their slopes are within about $10\%$ of the speed of light. We expect them to reach the speed of light in the limit $\bk \to \infty$.

\FIGURE[t]{\includegraphics[width=9cm]{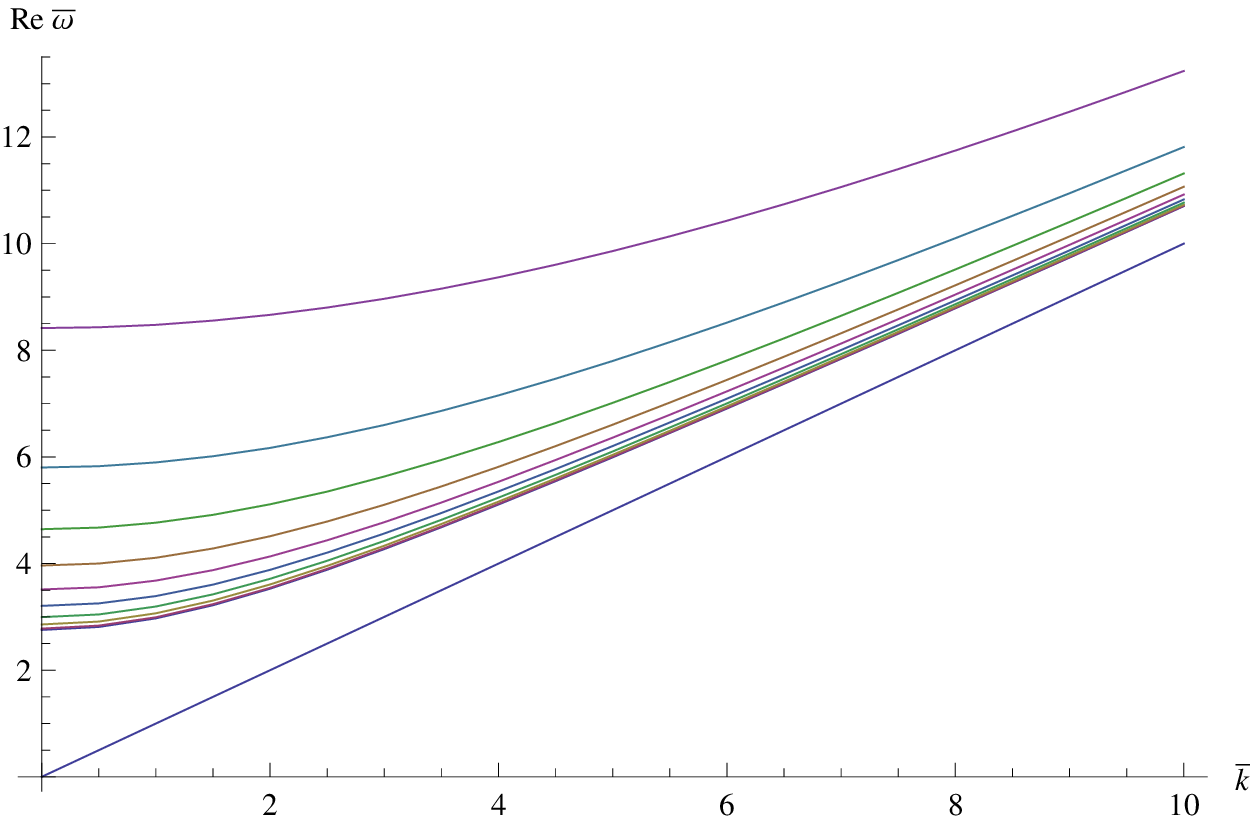}
\caption{(Color online) The real part of the first quasi-normal frequency of $\Phi(\brho)$ for $m=0$ (the point $n=1$ in figure~\ref{fig:Phi0QNMs}) as a function of $\bk$, for $\veps = M/\mu \in (0,0.9)$ in steps of $0.1$ from bottom to top. The solid blue line intersecting the origin is the lightcone, Re$(\bomg) = \bk$. We see that as $\bk$ increases the modes approach the lightcone. Numerically we find that at $\bk=10$ the asymptotic velocities are typically within about $10\%$ of the speed of light.}
\label{fig:asymvel}}

We also studied the QNM's of $\Phi(\brho)$ with $m=1$. The results were qualitatively similar to those for $m=0$, so we will not present them. In particular, none of the modes became unstable for the values of $\veps$ and $\bk$ we explored.

Our results for the QNM's of $\Phi^+(\brho)$ with $m=1$ are very similar to those for $\Phi(\brho)$ with $m=0$, although numerically the QNM's were more difficult to locate so we only present results for the first, in figure~\ref{fig:PhipQNMs}. Clearly the first QNM never crosses into the lower half of the $\bomg$ plane and hence never becomes unstable. On general grounds, we thus do not expect the higher $m=1$ QNM's, or the QNM's of $\Phi^+(\brho)$ with higher $m$, to become unstable.

\FIGURE[t]{\includegraphics[width=9cm]{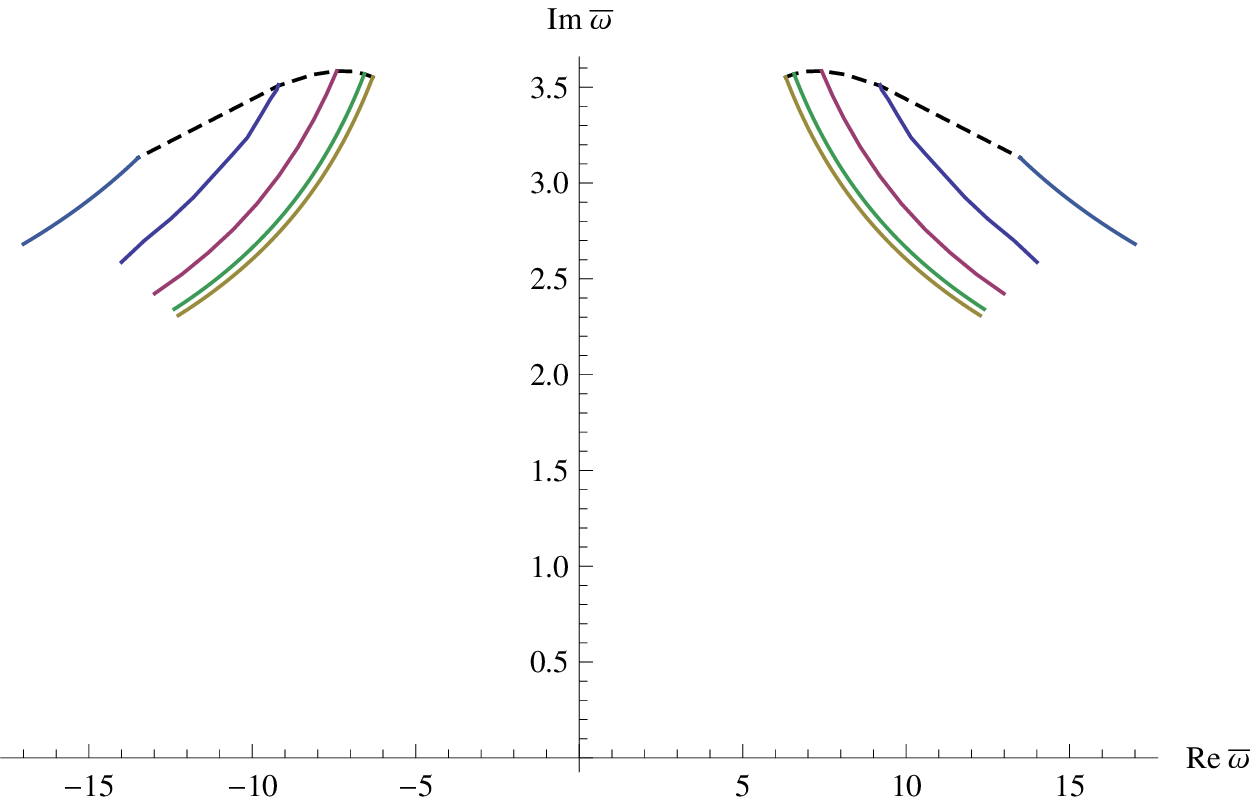}
\caption{(Color online) The complex $\bomg$ plane, where the curves show the motion of the first QNM of $\Phi^+(\brho)$ with $m=1$ for $\veps=(0,0.2,0.4,0.6,0.8)$ and $\bk \in (0,10)$. Our labeling conventions are the same as in figure~\ref{fig:Phi0QNMs}. The first QNM never crosses the real axis for the values of $\veps$ and $\bk$ that we consider, and hence never becomes unstable. The higher QNM's are more difficult to locate numerically. Given that the lowest mode never becomes unstable, and the higher modes are farther from the real $\bomg$ axis, we do not expect them to become unstable either, for our ranges of $\veps$ and $\bk$.}
\label{fig:PhipQNMs}}

Our results for the first three QNM's of $\Phi^-(\brho)$ with $m=1$ appear in figure~\ref{fig:PhiminusQNMs}. The results are qualitatively similar to those for the $\Phi(\brho)$ QNM's. In particular, none of the modes become unstable in the range of momenta we studied. The limiting velocities are qualitatively similar to those of the $\Phi(\brho)$, $m=0$ QNM's (see figure~\ref{fig:asymvel}).

\FIGURE[t]{\includegraphics[width=9cm]{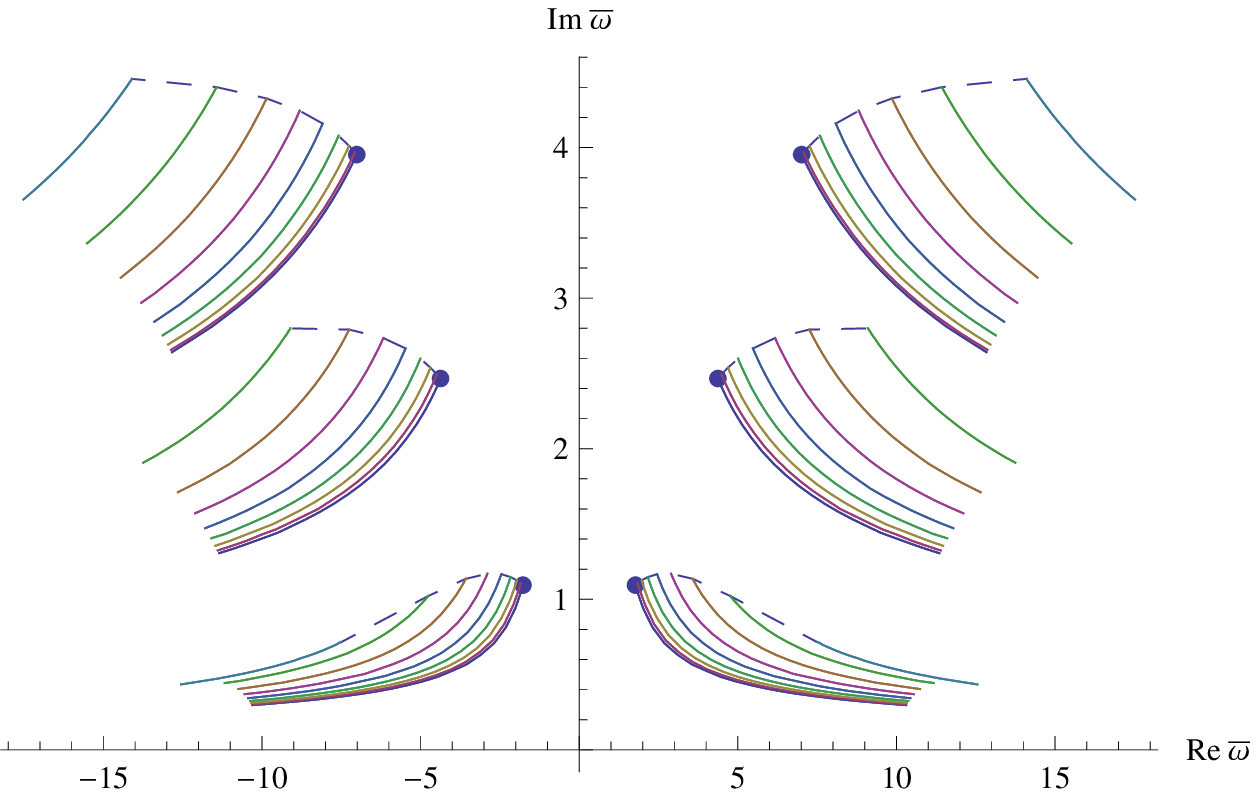}
\caption{(Color online) The complex $\bomg$ plane, where the curves show the motion of the first three QNM's of $\Phi^-(\brho)$ with $m=1$ as we vary $\veps$ and $\bk$. The lines have the same interpretation as those in figure~\ref{fig:Phi0QNMs}, although in this case for the $n=1$ and $n=3$ modes we only went up to $\veps=0.8$ and for the $n=2$ mode we only went up to $\veps=0.7$. We considered $\bk \in (0,10)$. The behavior of the QNM's is similar to that of the $\Phi(\brho)$ modes shown in figure~\ref{fig:Phi0QNMs}.}
\label{fig:PhiminusQNMs}}

A special QNM appears for $\Phi^-(\brho)$, qualitatively different from the others in that it sits precisely on the imaginary axis at small $\bk$. The motion of this mode as we increase $\bk$ is illustrated in figure~\ref{fig:hydromode}. For sufficiently small $\bk$, we have found via numerical fitting that the dispersion relation is approximately
\be
\label{eq:diffdef}
\bomg = i \frac{D(\veps)}{\sqrt{1-\veps^2}}\bk^2+ f(\veps) \bk^3 + {\cal O}(\bk^4),
\ee
for functions $D(\veps)$ and $f(\veps)$ that we can determine numerically. The leading behavior is similar to that of a diffusion mode for a conserved current, with diffusion constant $\frac{D(\veps)}{\sqrt{1-\veps^2}}$ (in terms of our dimensionless frequency and momentum $\bomg$ and $\bk$). Notice, however, that $\Phi^-(\brho)$ with $m=1$ is dual to the \textit{scalar} operator in eq.~\eqref{eq:oidef} \cite{Kruczenski:2003be}. A similar mode, with the same leading-order dispersion relation, also appears among the QNM's of $\Phi^-(\brho)$ with $m=2,3$. We plot the value of $\frac{D(\veps)}{\sqrt{1-\veps^2}}$ as a function of $\veps \in (0,1)$ for the three cases $m=1,2,3$ in figure~\ref{fig:diffusion}.

We discuss the physical interpretation of this ``diffusion mode'' in the next section.

\FIGURE[t]{\includegraphics[width=9cm]{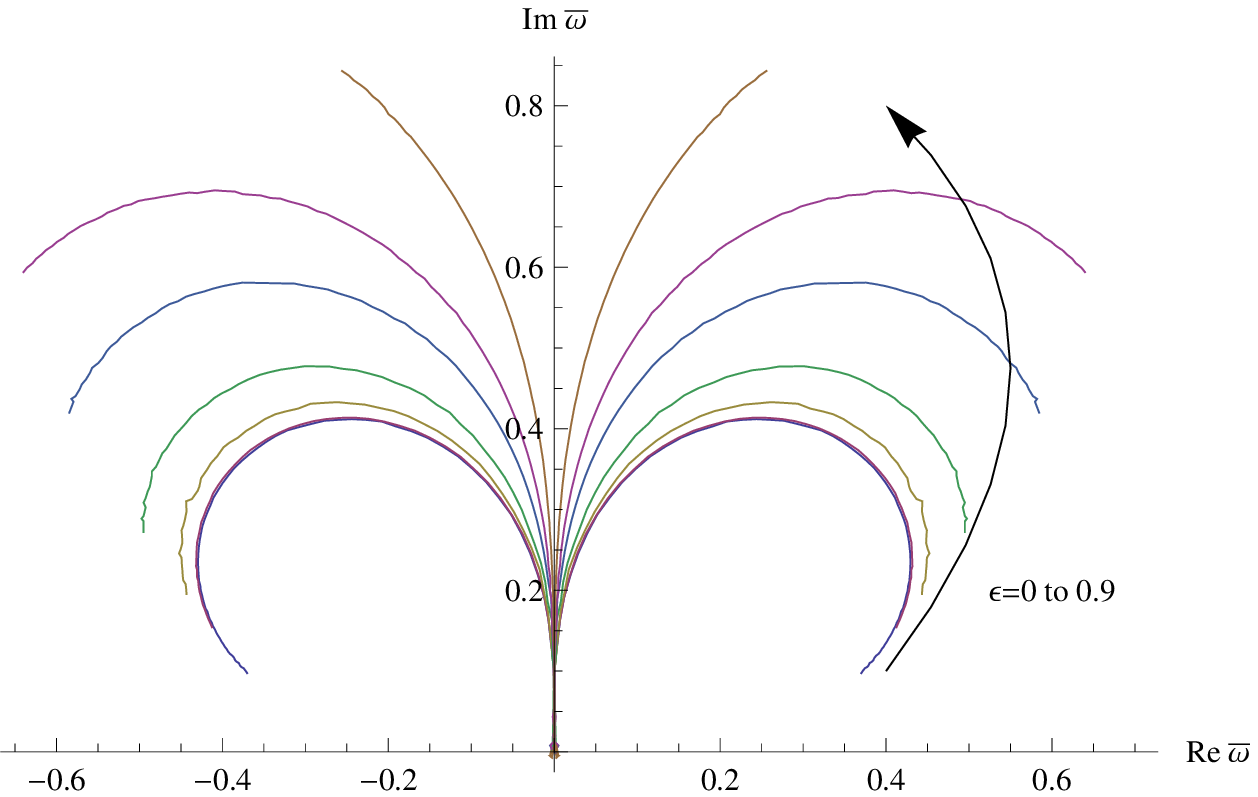}
\caption{(Color online) The complex $\bomg$ plane, where each line indicates the motion, as a function of momentum $\bk \in (0,9)$, of the $\Phi^-(\brho)$, $m=1$ QNM that starts on the imaginary axis at small $\bk$, our ``R-spin diffusion'' mode. The different lines are for $\veps=M/\mu=(0,0.1,0.3,0.5,0.7,0.8,0.9)$.}
\label{fig:hydromode}}

\FIGURE[t]{\includegraphics[width=9cm]{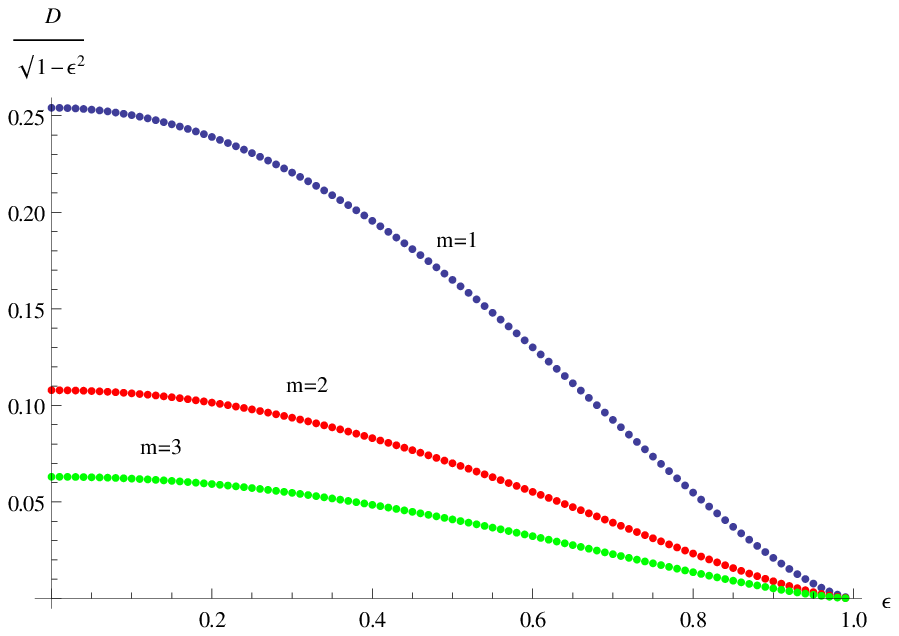}
\caption{(Color online) The ``diffusion constant'' $\frac{D(\veps)}{\sqrt{1-\veps^2}}$ defined in eq.~\eqref{eq:diffdef}, coming from the $\Phi^-(\brho)$, $m=1,2,3$ QNM's that lie on the imaginary axis at small momentum $\bk$, versus $\veps = M/\mu$.}
\label{fig:diffusion}}

\section{Discussion and Conclusions}\label{ss:conclusions}

We have studied the stability of holographic quantum critical matter in a top-down model, a single probe D7-brane in $AdS_5 \times S^5$ with nontrivial worldvolume gauge field, dual to $\N=4$ SYM theory coupled to a single flavor of $\N=2$ hypermultiplet, in a state with zero temperature and finite baryon density. We computed the spectrum of fluctuations of D7-brane fields, dual to the spectrum of mesons. For the range of momenta that we studied, $\bk \in (0,50)$, we detected no instabilities.

Beyond that, we had three main results. The first was to show that all bosonic worldvolume fluctuations effectively ``see'' a near-horizon $AdS_2$, due to the nontrivial background worldvolume fields. The dual field theory statement is that in these finite-density states, some (0+1) dimensional CFT emerges at low energies. The second main result was our numerical procedure, the zig-zag method, which may be useful for any calculation of QNM's for a bulk differential operator with singular points. Our third main result was the detection of a purely imaginary mode, similar to the diffusive mode of a conserved current, in the two-point function of the operator in eq.~\eqref{eq:oidef}, a dimension-two scalar operator that is a vector of $SU(2)_R$. If we interpret $SU(2)_R$ as spin, in analogy with electronic systems at low energy, then the mode we found has a natural interpretation as the diffusion of the ``R-spin.'' For higher values of $m$, the dual operators form an $(m+2)$-dimensional representation of $SU(2)_R$. For all of $m=1,2,3$, we saw a diffusion mode (see figure~\ref{fig:diffusion}), and we generically expect such a mode for all $m$.

An important task for the future is to extend our analysis to $N_f >1$. The flavor symmetry is then $U(1)_B \times SU(N_f)$, so we have the option of introducing a chemical potential for each Cartan generator of $SU(N_f)$, which in QCD language would be isospin chemical potentials. Here we expect that at zero temperature the ground state should in fact be a p-wave superfluid, representing a condensate of vector mesons \cite{Ammon:2008fc,Basu:2008bh,Ammon:2009fe,Erdmenger:2011hp}. Some natural questions then are whether any scale invariance emerges at low energies, \textit{i.e.} a (0+1)-dimensional CFT, and whether an R-spin diffusion mode appears.\footnote{For $N_f=2$ and at finite temperature, an isospin chemical potential, if represented in the bulk by a constant gauge field on the D7-branes, produces a instability towards squark condensation \cite{Apreda:2005yz}. Whether the same instability appears for a radially-varying gauge field remains an open question.}

Another important task for the future is to extend our analysis to other types of probe D-branes. An especially attractive example is a probe D5-brane extended along $AdS_4 \times S^2$ inside $AdS_5 \times S^5$ \cite{Karch:2000gx,DeWolfe:2001pq,Erdmenger:2002ex}, because an exact solution for the worldvolume fields, similar to the solution in eq.~\eqref{KO}, is known in that case \cite{Karch:2007br}. Here the dual flavor fields are restricted to a (2+1)-dimensional defect. The spectrum includes a zero sound mode \cite{Karch:2008fa}, and at very low energies we expect to find an emergent (0+1)-dimensional CFT \cite{Jensen:2010ga}. The spectrum of fluctuations will obviously be different from that of the D7-brane, and instabilities could appear. In particular, an instability involving a violation of the $AdS_2$ BF bound may occur, as happens for the D5-brane with an external $U(1)_B$ magnetic field \cite{Jensen:2010ga,Evans:2010hi}.

Our results add to the list of properties characterizing whatever microscopic state the solution in eq.~\eqref{KO} represents. That state has a low-temperature heat capacity $c_V \propto T^6$. The low-energy spectrum includes not only the zero sound mode but also the R-spin diffusion modes. At the lowest energies, some (0+1)-dimensional CFT emerges which controls the low-frequency form of all correlation functions. Despite having a nonzero entropy at zero temperature, the state is stable against small perturbations.

What effective theory replaces hydrodynamics at low temperatures? That theory must include not only the zero sound mode but also the R-spin diffusion modes. Can a Bose-Einstein condensate of strongly-coupled squarks, or some other more exotic microscopic physics, give rise to the properties listed above? In particular, what is the microscopic origin of the (0+1)-dimensional CFT? We plan to study these and related questions in the future.

\section*{Acknowledgements}

We would like to thank A.~Cherman, V.~Filev, S.~Hartnoll, M.~Kaminski, A.~Karch, P.~Kerner, K.-Y.~Kim, P.~Kumar, K.~Landsteiner, J.~Sonner, D.~Tong, and W.~Feng for helpful discussions. This work was supported by the Cluster of Excellence ``Origin and Structure of the Universe.'' The work of M.A. was supported by National Science Foundation grant PHY-07-57702. The work of S.L. was supported by the Alexander von Humboldt Foundation. The work of A.O'B. was supported by the European Research Council grant ``Properties and Applications of the Gauge/Gravity Correspondence.'' The work of J.S. was supported by the European Union through a Marie Curie Fellowship.

\appendix

\section*{Appendix: the Zero Sound Mode}\label{zerosound}

In this appendix we re-derive the results of refs.~\cite{Karch:2008fa,Kulaxizi:2008kv,Kim:2008bv} for the dispersion relation of the zero sound mode, which among other things provides a consistency check for our fluctuation equations of motion, eqs.~\eqref{EOM_ex} to \eqref{eome4}.

In the field theory, the zero sound mode appears in the $U(1)_B$ current two-point functions. These operators are singlets of the $SO(4)$ global symmetry. They correspond to fluctuations $\delta A_t$ and $\delta A_3$ that are zero modes on the $S^3$, that is, fluctuations with $m=0$. When the flavor mass is nonzero, so that the D7-brane embedding is non-trivial, these fluctuations couple to the $m=0$ mode of the fluctuation $\chi$ as well \cite{Kulaxizi:2008kv}. This suggests that the ansatz in our eq.~\eqref{eq:ansatz1} should include the zero sound mode, however that is not the case. When $m=0$ in that ansatz, eq.~\eqref{eome4} becomes trivial since the derivatives acting in $S^3$ directions vanish. Eq.~\eqref{eome4} is essentially a constraint equation that comes from setting $\delta A_i=0$. When a constraint is trivially satisfied, we expect to find an additional physical degree of freedom. In other words, an $m=0$ fluctuation distinct from that of eq.~\eqref{eq:ansatz1} should appear.

We thus consider an ansatz in which only the fluctuations $\chi$, $\delta A_t$, $\delta A_3$ are nonzero, all are zero-modes on the $S^3$, and all are plane waves of the form $e^{i \omg - i k x_3}$ in Minkowski directions. In this appendix, for all functions we will suppress all dependence on the coordinates $\rho$, $t$, and $x_3$ unless otherwise noted.

We will work in terms of two functions, $f_1$ and $f_2$, defined as
\be
\label{eq:phipsidefs}
f_1 = \delta A_t - \veps \chi, \qquad f_2 = \omg\eta^{\rho t}y'\chi+\omg\eta^{tt}/\eta^{xx}\dlt A_t-k\dlt A_3.
\ee
In terms of these, the constraint equation, eq.~\eqref{eomc}, becomes
\be
\del_\rho f_2=\omg\del_\rho(\eta^{tt}/\eta^{xx})(f_1).
\ee
The relevant equations of motion, eqs.~\eqref{eome1} and \eqref{eome2}, may be written as
\be
\label{eom1}
&&\del_a(\sqrt{-\eta}\eta^{ab}_S\eta^{xx}(1-\eta^{\rho\rho}\eta^{xx}y'{}^2)\del_b\chi)+\del_a(\sqrt{-\eta}\eta^{ab}_S\eta^{xx}\eta^{\rho t}y'\del_b\dlt A_t)+\omg\sqrt{-\eta}\eta^{xx}{}^2\eta^{\rho t}y'f_2=0, \no
\\
\label{eom2}
&&\del_a(\sqrt{-\eta}\eta^{ab}_S\eta^{xx}\eta^{\rho t}y'\del_b\chi)+\del_a(\sqrt{-\eta}\eta^{ab}_S\eta^{xx}(\eta^{tt}/\eta^{xx})\del_b\dlt A_t)+\omg\sqrt{-\eta}\eta^{tt}\eta^{xx}f_2=0.
\ee
Taking the linear combination of the above equations $-\frac{1}{\eta^{\rho\rho}\eta^{xx}} \left [ (\textrm{3}) + \veps (\textrm{4})\right]$, and making use of the identities
\be
\eta^{\rho t} y' = \frac{1}{\veps} \eta^{\rho \rho} \eta^{xx} y'{}^2 = - \veps \left(1+ \frac{\eta^{tt}}{\eta^{xx}}\right) = - \frac{1}{\veps} \left( \eta^{\rho\rho}\eta^{xx} + \frac{\eta^{tt}}{\eta^{xx}}\right),
\ee
we obtain $f_1$'s equation of motion,
\be\label{dynamical}
(\veps^2-1)\del_\rho(\eta^{tt}/\eta^{xx})\sqrt{-\eta}\del_\rho f_1+\del_a(\sqrt{-\eta}\eta^{ab}_S\eta^{xx}\del_bf_1)+\omg\sqrt{-\eta}\eta^{xx}{}^2f_2=0.
\ee
Crucially, notice that our fields $f_1$ and $f_2$ are not gauge-invariant, therefore we must also consider pure-gauge solutions \cite{Policastro:2002tn}. The advantage of our approach is that we only have one dynamical equation, eq.~\eqref{dynamical}, in contrast to the approach of ref.~\cite{Kulaxizi:2008kv}.

Notice that when the flavor mass $M=0$, so that $\veps=0$, $\chi$ disappears from eq.~\eqref{eq:phipsidefs}. As explained in section~\ref{ss:fluctuations}, that occurs because the chiral $U(1)_R$ symmetry is restored when $M=0$, and $\chi$ is charged under $U(1)_R$ but $\delta A_t$ and $\delta A_3$ are not. In the $M=0$ limit, we can identify the above fluctuations of the gauge field, along with $\delta A_1$ and $\delta A_2$, as the finite-density generalization of the type II fluctuations of ref.~\cite{Kruczenski:2003be}, with $m=0$.

To find the zero sound dispersion relation we follow refs.~\cite{Karch:2008fa,Kulaxizi:2008kv}: we solve the equations in two different limits and then match the solutions in a regime where the limits overlap. More precisely, we first take a near-horizon limit of the equation, $\rho \to 0$, and then expand the solution for low frequency and momentum, meaning $\omega /\rho \ll 1$ and $k /\rho \ll 1$ with $\omg/k$ fixed. We next take the low-frequency and low-momentum limit of the equation, solve that for all $\rho$, and then expand the solution near the horizon to perform the matching.

In the near-horizon limit, the in-going solutions are, with $\Omg=\omg\sqrt{1-\veps^2}$,
\be
\label{eq:nearhorizonsolszs}
f_1=c_1\, e^{-\frac{i\Omg}{\rho}}\, \rho^7(1+{\cal O}(\rho)), \qquad f_2=c_2\, e^{-\frac{i\Omg}{\rho}}\, \rho^2(1+{\cal O}(\rho)),
\ee
where the equations of motion dictate that the normalization constants $c_1$ and $c_2$ are related as
\be
c_2 = -\frac{6 i}{(1-\veps^2)^{1/2} \veps^2 \bar{N}^2} \, c_1,
\ee
where $\bar{N}$ is defined above eq.~\eqref{eq:rescaledvariables}. We expect to find such a relation, since we have only one dynamical equation, and hence only one integration constant.

In the low-frequency and low-momentum limit, eq.~\eqref{dynamical} becomes
\be
\label{eq:lowfreqphieqzs}
(\veps^2-1)\del_\rho(\eta^{tt}/\eta^{xx})\sqrt{-\eta}\del_\rho f_1+\del_\rho(\sqrt{-\eta}\eta^{\rho\rho}_S\eta^{xx}\del_\rho f_1)=0,
\ee
for which the solution is, in terms of the dimensionless radial coordinate $\brho$ defined in eq.~\eqref{eq:rescaledvariables},
\be
\label{phizs}
f_1(\rho)=C\int^{\brho}_0\frac{\brho'{}^6}{(\brho'{}^6+1)^{3/2}}d\brho',
\ee
where $C$ is an undetermined constant. Notice that eq.~\eqref{eq:lowfreqphieqzs} depends only on $\partial_{\rho} f_1$, so we can add any constant to $f_1$ to obtain a new solution. We have fixed the constant by demanding that the near-horizon limit of the solution match onto the low-frequency limit of the solution in eq.~\eqref{eq:nearhorizonsolszs}, that is, we demand that the two solutions agree in the regime where the two limits overlap, as advertised. The integral in eq.~\eqref{phizs} describes an incomplete Beta function, although we will not need that representation here. Integrating the constraint equation gives $f_2$,
\be
\label{psizs}
f_2(\rho)=\omg\int^{\brho}_0\frac{6}{\brho'{}^7(1-\veps^2)}f_1(\brho')d\brho'.
\ee
$f_1$ and $f_2$ approach constant values at the $AdS_5$ boundary,
\be
\label{physical}
\lim_{\rho \to \infty}f_1=C\frac{\Gamma(\frac{7}{6})\Gamma(\frac{4}{3})}{\sqrt{\pi}}, \qquad \lim_{\rho \to \infty}f_2=C\frac{2\omg}{1-\veps^2}\frac{\Gamma(\frac{7}{6})\Gamma(\frac{4}{3})}{\sqrt{\pi}}.
\ee

As emphasized earlier, we must also consider pure gauge solutions since our fields are gauge-dependent, and change under gauge transformations $\dlt A_b\rightarrow \dlt A_b+\del_b\alpha$. To preserve our gauge choice $\delta A_{\rho}=0$, $\alpha$ must be of the plane-wave form $e^{i\omg t-ik x_3}$, with no $\rho$ dependence. The pure-gauge solution is thus
\be
\chi=0, \quad \delta A_t = i \omg \, e^{i\omg t-ik x_3}, \quad \delta A_3 = -ik \, e^{i\omg t-ik x_3}.
\ee
Recalling the definitions of $f_1$ and $f_2$ in eq.~\eqref{eq:phipsidefs}, for the pure-gauge solution we find near the $AdS_5$ boundary (where $\eta^{tt}/\eta^{xx} \to -1$ in eq.~\eqref{eq:phipsidefs})
\be
\label{pure}
\lim_{\rho \to \infty}f_1=i\omg, \qquad \lim_{\rho \to \infty}f_2=-i(\omg^2-k^2).
\ee

The leading behavior of $f_1$ and $f_2$ near the boundary will be a linear combination of eqs.~\eqref{physical} and \eqref{pure}. These leading constants act as sources for the dual operators, which roughly speaking will vanish when $\omg$ is a QNM. More precisely, when $\omg$ is a QNM, the following condition holds:
\be
\det \left ( \begin{array}{cc} C\frac{\Gamma(\frac{7}{6})\Gamma(\frac{4}{3})}{\sqrt{\pi}} & C\frac{2\omg}{1-\veps^2}\frac{\Gamma(\frac{7}{6})\Gamma(\frac{4}{3})}{\sqrt{\pi}} \\ i \omg & -i(\omg^2-k^2) \end{array} \right)= i C \frac{\Gamma(\frac{7}{6})\Gamma(\frac{4}{3})}{\sqrt{\pi}} \left(k^2 - \omg^2 - \frac{2 \omg^2}{1-\veps^2}\right)= 0,
\ee
which gives us the leading term in the dispersion relation,
\be\label{zs_disp}
\frac{\omg^2}{k^2}=\frac{1-\veps^2}{3-\veps^2} = \frac{\mu^2 - M^2}{3\mu^2 - M^2},
\ee
which agrees with the results for the speed of zero sound squared in ref.~\cite{Kulaxizi:2008kv} when $M\neq0$ and in ref.~\cite{Karch:2008fa} (and eq.~\eqref{eq:zsdispersion}) when $M=0$.

We should mention that, like all of the other bosonic worldvolume fluctuations, the fluctuations giving rise to zero sound ``see'' an effective $AdS_2$ near-horizon region \cite{Nickel:2010pr}. That is clear from eq.~\eqref{eq:nearhorizonsolszs}, since the $e^{- i\Omega/\rho}$ factor is a solution of eq.~\eqref{eq:nearhorizonphiscalarrescaled}, indicating that after removing a power of the radial coordinate, $\rho^7$ in $f_1$'s case and $\rho^2$ in $f_2$'s case, in the near-horizon region the fluctuations obey an equation of motion identical to a massless scalar in $AdS_2$.

\bibliographystyle{JHEP}
\bibliography{d7zerotemp}

\end{document}